\documentclass{aa}  

\usepackage{graphicx}
\usepackage{txfonts}
\usepackage{natbib}
\bibpunct{(}{)}{;}{a}{}{,} 
\usepackage[]{hyperref}
\hypersetup{colorlinks,
citecolor=blue
}
\newcommand{\fluxunit}{$\textrm{erg}\,\textrm{s}^{-1}\,\textrm{cm}^{-2}$}

\begin{document} 

\title{An H$\alpha$ view of galaxy build-up in the first 2 Gyr: luminosity functions at z$\sim4-6.5$ from NIRCam/grism spectroscopy}
\titlerunning{An H$\alpha$ view of galaxy build-up in the first 2 Gyr}

   \author{Alba Covelo-Paz\inst{1},
          Emma Giovinazzo\inst{1},
          Pascal A. Oesch\inst{1,2,3},
          Romain A. Meyer\inst{1},
          Andrea Weibel\inst{1},
          Gabriel Brammer\inst{2,3},
          Yoshinobu Fudamoto\inst{4},
          Josephine Kerutt\inst{5},
          Jamie Lin\inst{6},
          Jasleen Matharu\inst{2,3},
          Rohan P. Naidu\inst{7,8},
          Anna Velichko\inst{1,9},
          Victoria Bollo\inst{10},
          Rychard Bouwens\inst{11},
          John Chisholm\inst{12},
          Garth D. Illingworth\inst{13},
          Ivan Kramarenko\inst{14},
          Daniel Magee\inst{13},
          Michael Maseda\inst{15},
          Jorryt Matthee\inst{14},
          Erica Nelson\inst{16},
          Naveen Reddy\inst{17},
          Daniel Schaerer\inst{1},
          Mauro Stefanon\inst{18,19},
          Mengyuan Xiao\inst{1}
          }

   \institute{Department of Astronomy, University of Geneva, Chemin Pegasi 51, 1290 Versoix, Switzerland\\ 
   \email{alba.covelopaz@unige.ch}
   \and
   Cosmic Dawn Center (DAWN), Copenhagen, Denmark
   \and
   Niels Bohr Insitute, University of Copenhagen, Jagtvej 128, 2200 Copenhagen, Denmark
   \and
   Center for Frontier Science, Chiba University, 1-33 Yayoi-cho, Inage-ku, Chiba 263-8522, Japan
   \and
   Kapteyn Astronomical Institute, University of Groningen, P.O. Box 800, 9700 AV Groningen, The Netherlands
   \and
   Department of Physics and Astronomy, Tufts University, 574 Boston Avenue, Medford, MA 02155, USA
   \and
   MIT Kavli Institute for Astrophysics and Space Research, 77 Massachusetts Ave., Cambridge, MA 02139, USA
   \and
   NASA Hubble Fellow
   \and
   Institute of Astronomy, Kharkiv National University, 4 Svobody Sq., Kharkiv, 61022, Ukraine
   \and
   European Southern Observatory, Karl-Schwarzschildstrasse 2, D-85748 Garching bei München, Germany
   \and
   Leiden Observatory, Leiden University, PO Box 9513, 2300 RA Leiden, the Netherlands
   \and
   Department of Astronomy, The University of Texas at Austin, 2515 Speedway, Stop C1400, Austin, TX 78712-1205, USA
   \and
   Department of Astronomy and Astrophysics, University of California, Santa Cruz, CA 95064, USA
   \and
   Institute of Science and Technology Austria (ISTA), Am Campus 1, 3400 Klosterneuburg, Austria
   \and
   Department of Astronomy, University of Wisconsin–Madison, 475 N. Charter St., Madison, WI 53706 USA
   \and
   Department for Astrophysical and Planetary Science, University of Colorado, Boulder, CO 80309, USA
   \and
   Department of Physics and Astronomy, University of California, Riverside, 900 University Avenue, Riverside, CA 92521, USA
   \and
   Departament d'Astronomia i Astrof\`isica, Universitat de Val\`encia, C. Dr. Moliner 50, E-46100 Burjassot, Val\`encia,  Spain
   \and
   Unidad Asociada CSIC "Grupo de Astrof\'isica Extragal\'actica y Cosmolog\'ia" (Instituto de F\'isica de Cantabria - Universitat de Val\`encia), Spain
}
\authorrunning{A. Covelo-Paz et al.}
   \date{Received XX; accepted YY}

 
\abstract{The H$\alpha$ nebular emission line is an optimal tracer for recent star formation in galaxies. With the advent of \emph{JWST}, this line has recently become observable at $z>3$ for the first time.  We present a catalog of $1\,050$ H$\alpha$ emitters at $3.7<z<6.7$ in the GOODS fields obtained from a blind search in \emph{JWST} NIRCam/grism data. We make use of the FRESCO survey's 124 arcmin$^2$ of observations in GOODS-North and GOODS-South with the F444W filter, probing H$\alpha$ at $4.9<z<6.7$; and the CONGRESS survey's 62 arcmin$^2$ in GOODS-North with F356W, probing H$\alpha$ at $3.8<z<5.1$. We find an overdensity with 98 sources at $z\sim4.4$ in GOODS-N and confirm previously reported overdensities at $z\sim5.2$ in GOODS-N and at $z\sim5.4$ and $z\sim 5.9$ in GOODS-S. We compute the observed H$\alpha$ luminosity functions (LFs) in three bins centered at $z\sim4.45$, $5.30$, and $6.15$, which are the first such measurements at $z>3$ obtained based purely on spectroscopic data, robustly tracing galaxy star formation rates (SFRs) beyond the peak of the cosmic star formation history. We compare our results with theoretical predictions from three different simulations and find good agreement at $z\sim4-6$. The UV LFs of this spectroscopically-confirmed sample are in good agreement with pre-\emph{JWST} measurements obtained with photometrically-selected objects. Finally, we derive SFR functions and integrate these to compute the evolution of the cosmic star-formation rate densities across $z\sim4-6$, finding values in good agreement with recent UV estimates from Lyman-break galaxies, which imply a continuous decrease in SFR density by a factor of 3$\times$ over $z\sim4$ to $z\sim6$. Our work shows the power of NIRCam grism observations to efficiently provide new tests for early galaxy formation models based on emission line statistics.
}

   \keywords{Galaxies: evolution -- Galaxies: formation -- Galaxies: high-redshift -- Galaxies: luminosity function, mass function -- Galaxies: star formation}

   \maketitle
%
\section{Introduction}
Uncovering the assembly histories of galaxies throughout cosmic time is a primary objective of modern astrophysics. In the early universe, the star formation rate density (SFRD) increases with time up to a peak at $z\sim2-3$, at the so-called `cosmic noon' \citep{MD14}.
Probing the SFRD at high redshifts is essential to shed light on the process of early galaxy assembly. However, until recently the SFR estimates at $z>3$ almost exclusively relied on rest-frame UV continuum estimates, which involved an uncertain dust correction \citep[e.g.,][]{Wyder05,Schiminovich05,Dahlen07,ReddySteidel09,RobothamDriver11,Cucciati12,Bouwens12b,Bouwens12a,Schenker13,Bouwens+15}. 

At lower redshifts, the H$\alpha$ emission line is among the most commonly used estimators for galaxy star formation \citep[e.g.,][]{Erb06,Geach08,Forster09,Hayes10,Sobral+13}. This line is the result of hydrogen gas recombination in the presence of ionizing photons emitted by the most massive stars in a galaxy. These stars live for a short period of $<10$ Myr, making H$\alpha$ an optimal tracer for recent star formation in a galaxy \citep[e.g.,][]{Kennicutt+98}. Moreover, H$\alpha$ is less affected by dust attenuation than the UV continuum.
Multiple studies used H$\alpha$ measurements to determine star formation densities at various redshifts (e.g., \citealt{Stroe&Sobral15} at $0.19<z<0.23$, \citealt{Sobral+13} at $0.4<z<2.3$, \citealt{Terao+22} at $2.1<z<2.5$, \citealt{Bollo23} at $3.86<z<4.94$). 
With ground-based near-IR spectrographs such as MOSFIRE and KMOS, spectroscopic detections of H$\alpha$ became standard at $1.4<z<2.6$ \citep[e.g.,][]{Kashino+13,Steidel+14,Kriek+15}.

With the recent advent of \emph{JWST} \citep{Gardner+06,Rigby+23}, the H$\alpha$ line is now accessible for high-resolution spectroscopy and imaging at $z>2.5$, and some studies have already shown its capabilities \citep[e.g.,][]{Nelson23,Matharu24,Pirie+24}. This calls for a blind, luminosity-limited search of H$\alpha$ emitters to provide a stringent estimate of the star formation rates in a well-defined and complete sample. In this paper, we present the results of a blind search of H$\alpha$ emitters at $3.8<z<6.6$ in the GOODS fields. For this, we use the \emph{JWST} First Reionization Epoch Spectroscopically Complete
Observations (FRESCO; Cycle 1, \citealt{Oesch23}) and the Complete NIRCam Grism Redshift Survey (CONGRESS; Cycle 2, \citealt{Congress}) programs. FRESCO obtained NIRCam imaging and slitless spectroscopy with the F444W filter in an area of 62 arcmin$^2$ in GOODS-North and 62 arcmin$^2$ in GOODS-South. CONGRESS used the same configuration as FRESCO and covered the same 62 arcmin$^2$ in GOODS-North with the F356W filter.
We use this new catalog of spectroscopically-confirmed H$\alpha$ emitters to obtain the H$\alpha$ luminosity functions (LFs) at $z>2.5$ derived from spectroscopic data for the first time, and derive new cosmic SFRD measurements at redshifts $z\sim4.45$, $5.3$, and $6.15$. 

This paper is structured as follows. In Sect.~\ref{sec:data} we describe our data reduction to obtain the full catalog of H$\alpha$ emitters. In Sect.~\ref{sec:completeness} we detail the completeness characterization of this catalog. In Sect.~\ref{sec:results} we obtain the H$\alpha$ LFs, star formation rate functions, and SFRDs at $z\sim4.45,5.3,$ and $6.15$. In Sect. \ref{sec:discussion} we discuss our findings and compare them to previous studies at redshifts $z=0-10$.
Throughout this work, we adopt a concordance flat $\Lambda$CDM cosmology with $H_0=70\,\textrm{km}\,\textrm{s}^{-1}\,\textrm{Mpc}^{-1}$, $\Omega_m=0.3$, and $\Omega_\Lambda=0.7$. Magnitudes are listed in the AB system \citep{Oke&Gunn}. 

\section{Data}
\label{sec:data}
\subsection{Observations}
\label{sec:observations}
We use NIRCam/grism spectra with the F444W filter from FRESCO \citep{Oesch23} and the F356W filter from CONGRESS (Sun, Egami et al., in prep., \citealt{Congress}). Both surveys observed 62 arcmin$^2$ of the GOODS-North field, and their combined spectra cover the wavelength range $3.1-5.0$ $\mu$m. Moreover, FRESCO observed 62 arcmin$^2$ of the GOODS-South field, covering $3.8-5.0$ $\mu$m. The combination of FRESCO and CONGRESS data allows us to perform a blind search of H$\alpha$ emitters in the redshift range of $3.7<z<6.7$ in GOODS-North, and $4.9<z< 6.7$ in GOODS-South.

For our photometric catalogs in the two GOODS fields, we use all the publicly available HST \citep[e.g.][]{Giavalisco2004, Grogin2011, Koekemoer2011} and JWST \citep[e.g.][]{Eisenstein2023a, Eisenstein2023b, Williams2023} imaging in the MAST archive\footnote{This includes partial or full coverage by the ACS F435W, F606W, F775W, F814W, F850LP, and WFC3 F105W, F125W, F140W, and F160W filters, as well as the NIRCam F090W, F115W, F150W, F182M, F200W, F210M, F277W, F335M, F356W, F410M and F444W filters in both fields. The GOODS-S field has additional partial coverage by the NIRCam F162M, F250M, F300M, F430M, F460M and F480M, as well as the MIRI F560W and F770W filters.}. The basic procedure applied to derive photometric catalogs is described in \citet{Weibel2024}. To ensure a homogeneous depth for the detection image across the FRESCO survey footprint, we generate custom mosaics of the F444W and F210M images using imaging from FRESCO only, and excluding any available data from other surveys in these filters. These reductions are combined in an inverse-variance weighted stack to obtain a detection image with a typical $5\sigma$ depth of 28.5 mag in 0.16\arcsec radius circular apertures. We run \texttt{SourceExtractor} \citep{Bertin1996} in dual mode to measure fluxes through circular apertures with a radius of 0.16\arcsec in all available filters, including data from all available surveys. Prior to measuring fluxes, all images are matched to the point spread function (PSF) resolution in the F444W filter. We apply an aperture correction based on fluxes measured through Kron apertures on a PSF-matched version of the detection image and scale the fluxes to total by accounting for the encircled energy of the Kron ellipse on the F444W PSF.

\subsection{Grism Extractions}
\label{sec:grism}
The NIRCam/grism data are reduced and processed with the publicly available \texttt{grizli} code\footnote{\url{https://github.com/gbrammer/grizli}}, using the standard CRDS grism dispersion files from pmap 1123 made available in September 2023 and slightly modified sensitivity functions from \texttt{grizli}\footnote{available at \url{https://s3.amazonaws.com/grizli-v2/JWSTGrism/NircamSensitivity/index.html}} (see also, e.g., \citealt{Oesch23}, \citealt{Matharu24},  \citealt{Meyer+24}).

Given the limited wavelength range of the grism spectra, sources with H$\alpha$ emission lines often do not show any other strong line in the F444W filter (except for sources with additional F356W grism spectra in GOODS-N). Additionally, given the single dispersion angle of the FRESCO data,  the association of a given emission line with an individual galaxy along the dispersion direction is not always unique. It is therefore not straightforward to blindly search for H$\alpha$ emitters. Instead, we need to rely on initial redshift guesses, such as an existing spectroscopic redshift or a photometric redshift estimate to search for H$\alpha$ emission lines in galaxies. Therefore, we extract grism spectra only for sources, for which H$\alpha$ lines are expected to lie in the grism wavelength range with a non-zero probability. In particular, our extractions are based on several criteria:
(1) sources with \texttt{eazy} photometric redshifts at $z_\mathrm{phot}=4.85-6.7$ for the F444W FRESCO observations or $z_\mathrm{phot}=3.7-5.2$ for the F356W CONGRESS field and with photometric redshift uncertainties of less than 20\% and with $m_\mathrm{444}<28$ AB mag. 
(2) we additionally extract all sources with ancillary spectroscopic redshifts in the same ranges stated above.

We then run a line search with \texttt{grizli} over the 2.5 to 97.5 percentiles of the redshift probability density functions as measured with \texttt{eazy}. For sources with spectroscopic redshifts, we search within $0.02\times(1+z_\mathrm{spec})$ of the spectroscopic redshift, which corresponds to velocity offsets of up to $\sim$6000 km/s.

\subsection{Selection of H$\alpha$ emitters}
\label{sec:selection}

After the line search, we selected the sources with H$\alpha$ $\textrm{S/N}>3$ from the \texttt{grizli}-extracted catalog.  
Upon visual inspection, we found several multi-component sources presenting either very nearby neighbors or clumpy disks that were split into different components during the \texttt{SourceExtractor} run. As a result, several sources in our catalog were clumps from the same galaxy, which were assigned the same H$\alpha$ emission line. To avoid accounting for the same galaxy more than once, we adopted a similar procedure to \citet{Matthee+23}: we grouped sources with identical grism redshifts and an angular separation $<0.6$ arcsec, which corresponds to a physical distance of $\lesssim3.3-4.4$ kpc at the studied redshift range ($3,8<z<6.6$). We then re-extracted the spectra of the multi-component objects as one source with \texttt{grizli} and re-computed the photometry and object positions as flux-weighted centroids. This led to a sample of $1\,081$ sources in CONGRESS and 858 in FRESCO, where $\sim10\%$ of these are multi-component galaxies. 

Each source of this sample was then visually inspected by four team members using the tool \texttt{specvizitor}\footnote{\url{https://github.com/ivkram/specvizitor}}. Each individual inspector assigned a quality flag to every source, following the classification in \citet{Meyer+24}: a quality flag $q=3$ was assigned to sources with a clear H$\alpha$ line and a matching morphology between the direct image and the 2D-spectrum and line map; $q=2$ to sources with a lower S/N, but a clear morphology match; $q=1$ to unclear sources; $q=0$ to objects lacking an H$\alpha$ line; and $q=-1$ to contamination by neighboring objects. 

\begin{figure*}
    \centering
    \includegraphics[width=0.49\linewidth]{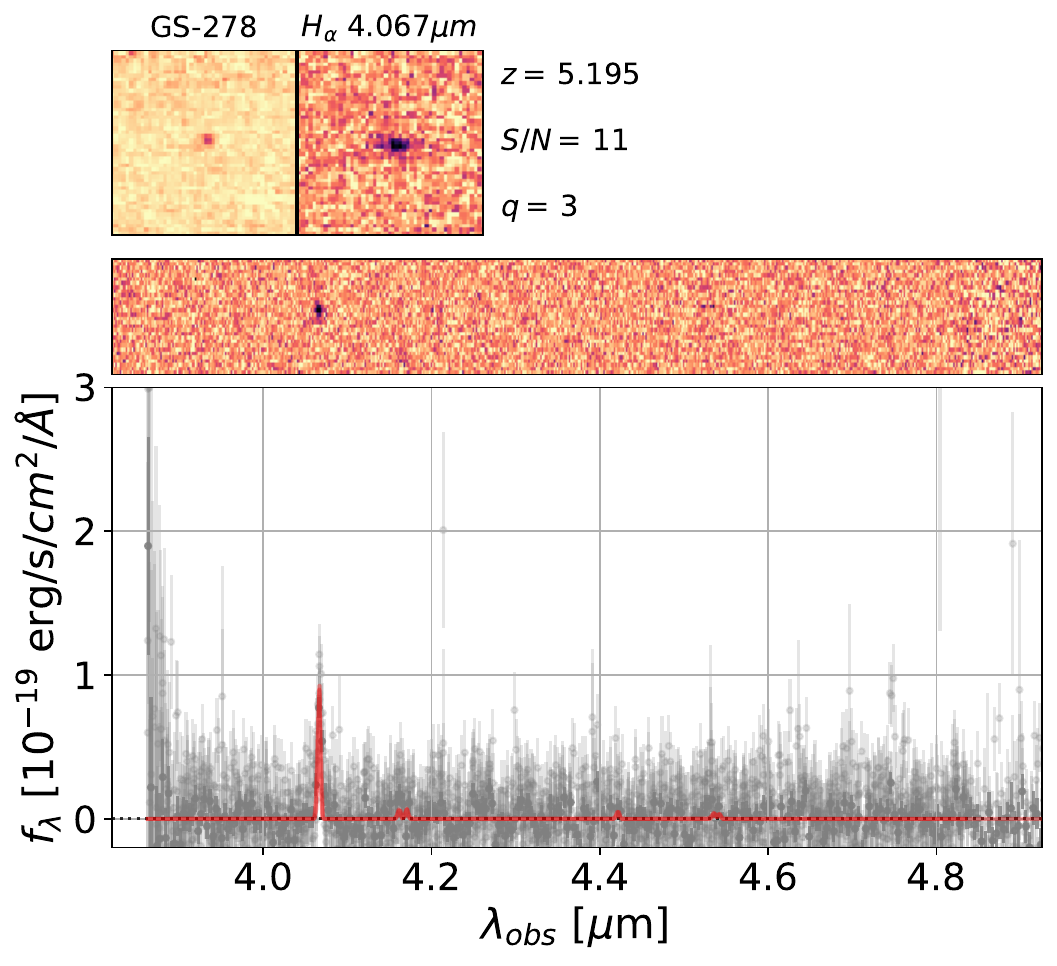}    \includegraphics[width=0.49\linewidth]{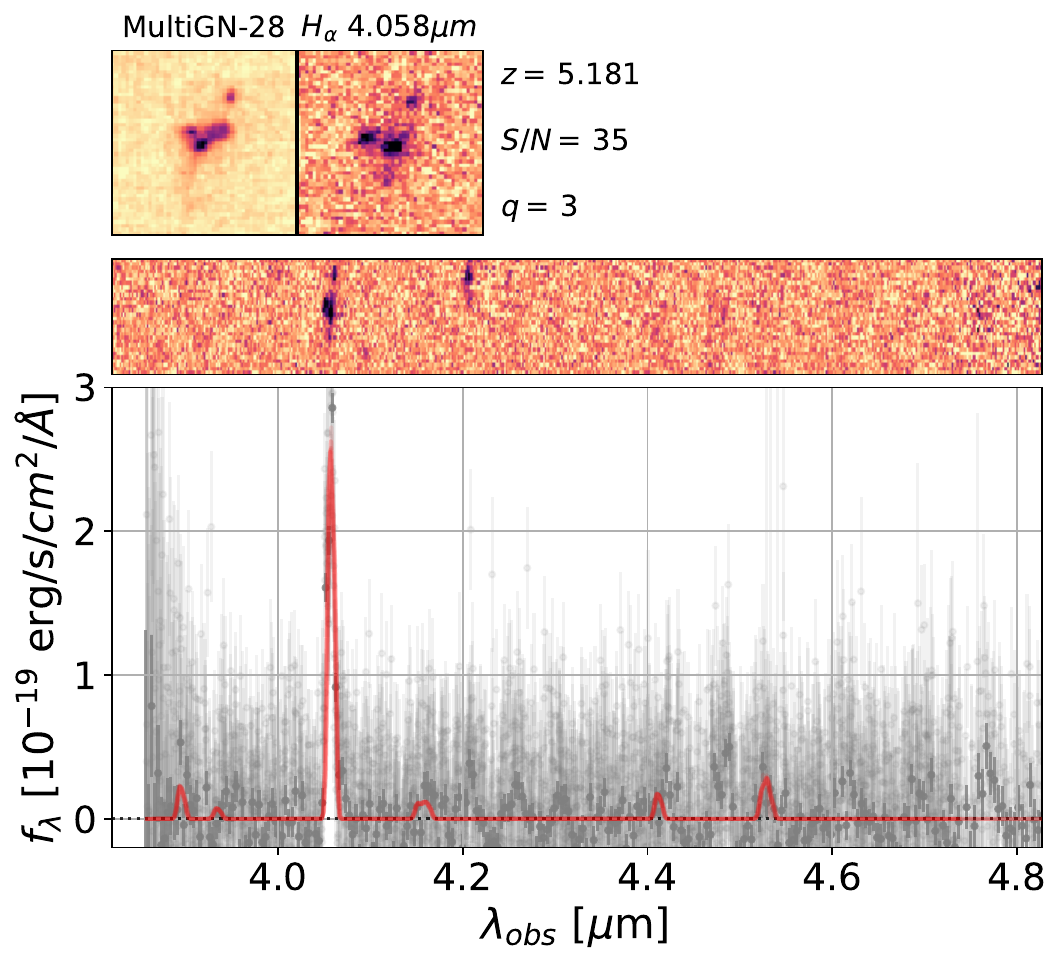}    \includegraphics[width=0.49\linewidth]{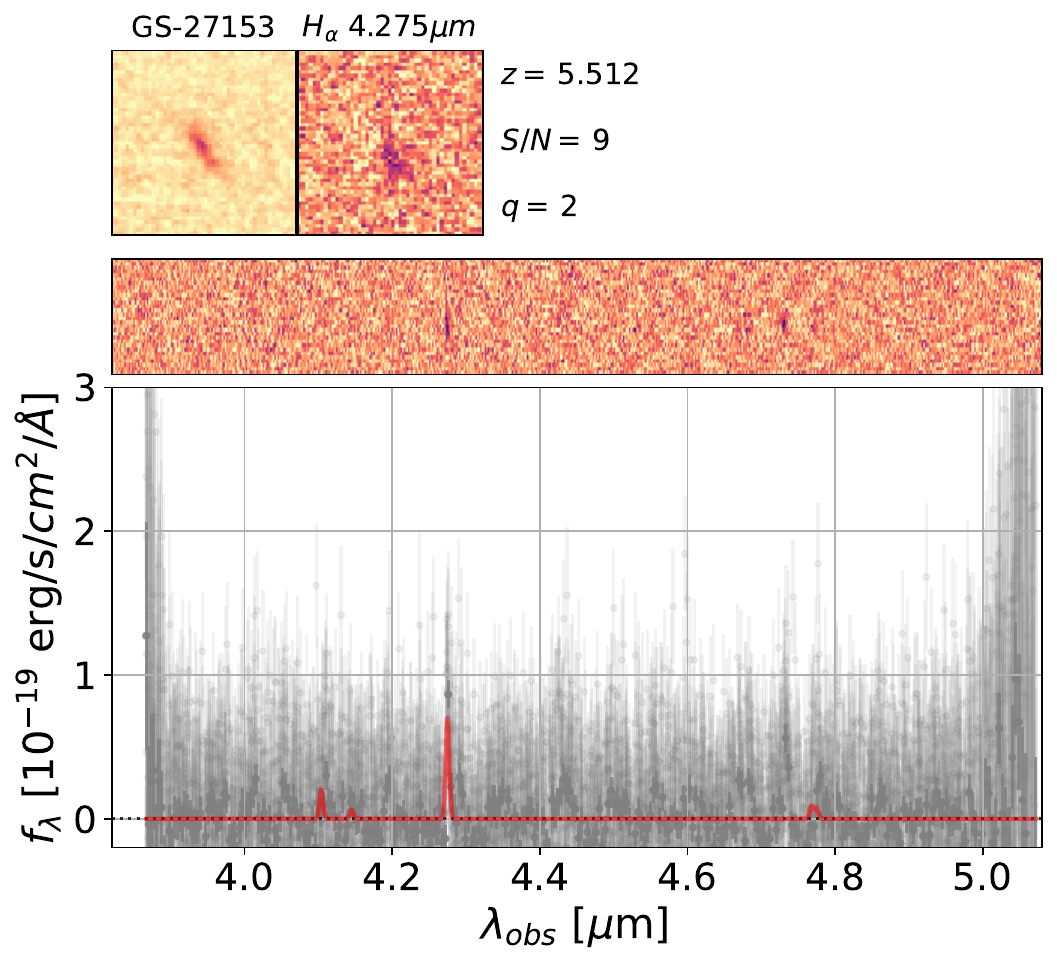}    \includegraphics[width=0.49\linewidth]{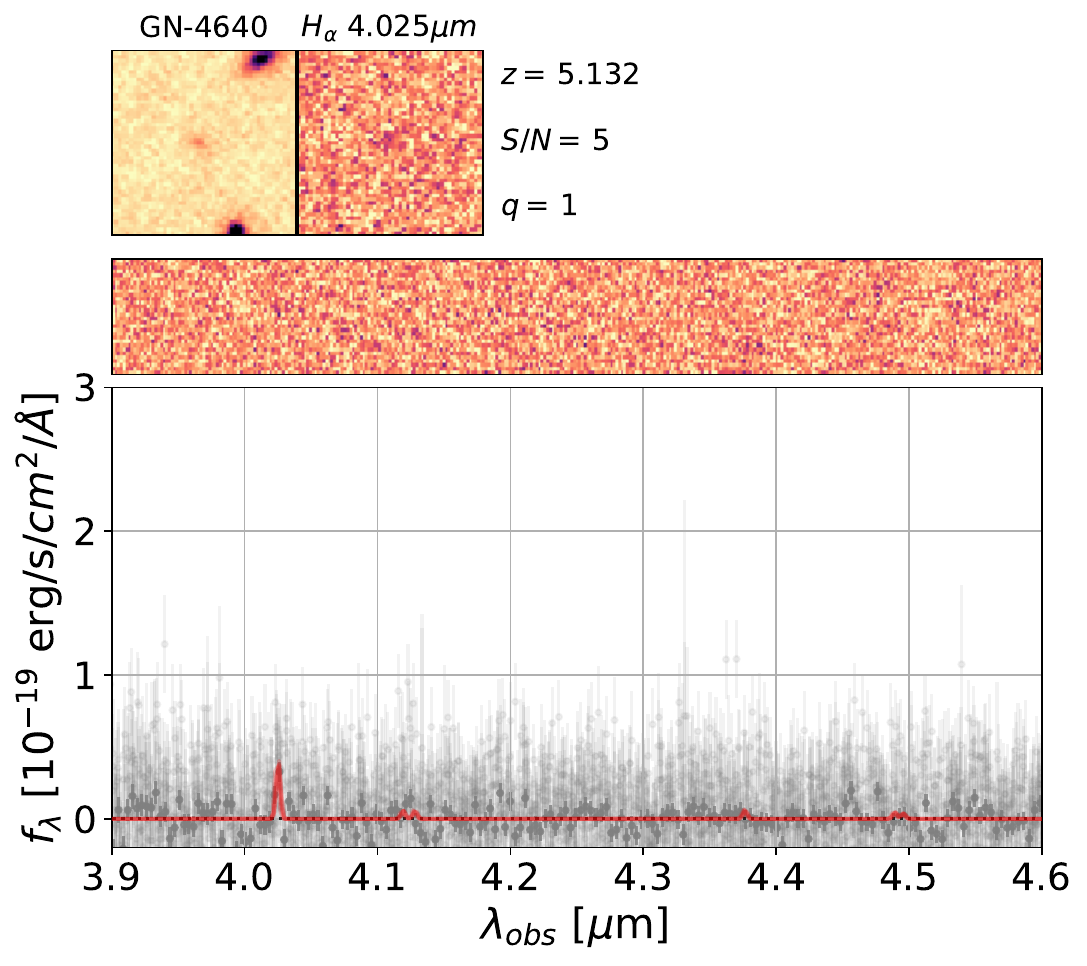}
    \caption{Direct F444W stamps, H$\alpha$ maps, 2D-spectra, and 1D-spectra of four sources in the FRESCO H$\alpha$ catalog with different quality flags. Top left: single-component galaxy with $q=3$. Top right: multi-component galaxy with $q=3$. Bottom left: single-component with $q=2$. Bottom right: single-component with $q=1$. For the analysis of this paper, we only include sources with $q\geq1.75$.}
    \label{specs}
     \end{figure*}

The FRESCO H$\alpha$ emitters in GOODS-North with $z>5.25$ also show [\ion{O}{iii}] $5\,008\AA$ emission in CONGRESS data. This enables us to further check the quality of the 179 FRESCO sources that have a valid CONGRESS spectrum within the H$\alpha$ range, and we find that 36 sources have an improved quality flag after inspecting CONGRESS, while only five have been downgraded. The final quality flag of each source is the average of the four inspector grades. Fig. \ref{specs} shows the extractions of four sources with different quality flags. 

For the final H$\alpha$ catalog we selected sources with an overall quality flag $q\geq 1.75$, meaning that a majority of the team believed these lines to be likely real. The result is a catalog with $1\,050$ H$\alpha$ emitters: 554 from CONGRESS (in GOODS-N) and 503 from FRESCO (318 in GOODS-N and 185 in GOODS-S), with seven sources in both surveys. Despite our original S/N cut of $\textrm{S/N}>3$ for the parent sample, all the sources flagged with $q\geq 1.75$, and thus in our final catalog, turned out to have $\textrm{S/N}>5$. 

The full H$\alpha$ catalog can be found at \url{https://github.com/astroalba/fresco}. An excerpt is shown in table \ref{tab:cat}. 

\subsection{Comparison to JADES NIRSpec Catalogs}

123 of the sources in our H$\alpha$ catalog are also present in JADES DR3 \citep{JADES_DR3}, which contains prism and grating redshifts for targeted sources in both GOODS fields. After comparing our grism redshifts to JADES prism redshifts, we found these to agree very well, within a scatter of $0.001\times(1+z)$ and no outliers. We also compare our H$\alpha$ fluxes to those of JADES DR3 in appendix \ref{sec:jades}. The grism and NIRSpec line flux measurements are broadly consistent albeit with a significant dispersion of up to 0.26 dex. While the grism data measures the full (spatially resolved) emission line without flux loss (in principle), the NIRSpec line flux measurements need to be corrected for slit losses. It is likely that this correction as well as different relative calibrations introduce this dispersion in flux measurements. Overall, this comparison is very reassuring. A more detailed comparison between the NIRSpec catalogs and the NIRCam grism line fluxes is deferred to our data release paper.

\subsection{Removal of broad-line AGN}
During the visual inspection process, we found 13 of our sources to show a significant broad component in their H$\alpha$ lines (FWHM$>$800 km\,s$^{-1}$) consistent with AGN activity. These sources are reported in appendix \ref{sec:LRDs}. Eight of them are in FRESCO and match those previously reported as Little Red Dots by \citet{LRDs}. The other five are found at lower redshifts in the CONGRESS data and are new discoveries. These sources are removed from our sample of LF measurements as we are only interested in star-forming galaxies in this work. We note, however, that our results do not change significantly with or without these sources.

\subsection{UV luminosities}
In order to compare our results with a different star formation rate tracer, we derive UV luminosities (and absolute magnitudes) using the fluxes in our photometric catalog. For each source, we determine which filters lie in the rest-frame UV wavelength range, i.e. $1\,400-2\,700\;\AA$. We fit these fluxes to $f_\nu\propto\lambda^{\beta-2}$ by doing a linear fit in $\textrm{log}-\textrm{log}$ space, and then interpolate the flux value at $1\,500\;\AA$. This $f_{1500;\textrm{UV}}$ is then converted to $L_\textrm{UV}$. 

\subsection{SED fitting and dust corrections}
\label{sec:sed}
To account for dust attenuation, we fit the photometry of our sources using the \texttt{Bagpipes} SED fitting code \citep{Bagpipes}. 
We adopt a star formation history consisting of a delayed tau model and a recent burst which we allow to occur between $0-10$ Myr before the time of observation. We use a \citep{Calzetti+00} dust attenuation curve. The redshift is fixed to the grism spectroscopic redshift, however, we do not include the grism emission line fluxes in the fit. We allow the visual attenuation to vary from $A_V=0-5$ mag, and the metallicity from $Z=0.1Z_\odot-Z_\odot$, using uniform priors. Based on the resulting visual attenuation, $A_V$, we further derive the dust attenuation at the  H$\alpha$ wavelength based on the \citet{Calzetti+00} curve, 
assuming that the nebular reddening is the same as the continuum reddening.

86 of the sources in our H$\alpha$ catalog have Balmer decrements measured by JADES NIRSpec spectroscopy as released in DR3 \citep{JADES_DR3}. JADES obtained H$\alpha$ and H$\beta$ fluxes for targeted sources in both GOODS fields using NIRSpec G140M/G235M/G395M grating and prism spectroscopy. This allowed us to compare the H$\alpha$ dust attenuation, $A_{\textrm{H}\alpha}$, computed by \texttt{Bagpipes} SED fitting to that measured directly from the Balmer decrement, using the same dust attenuation law from \citet{Calzetti+00}. We find that both methods provide $A_{\textrm{H}\alpha}$ results with large uncertainties; nevertheless, the median H$\alpha$ dust attenuation for the sample is consistent with both methods: $A_{\textrm{H}\alpha\textrm{,med}}=0.47$ mag. Throughout this paper, we thus use the H$\alpha$ dust attenuation values from \texttt{Bagpipes} derived for each source. 

\section{Completeness}
\label{sec:completeness}
To accurately compute luminosity number densities, we need to account for the completeness of our sample, which depends on the continuum as well as the line fluxes and the position of the sources in the field. Following each step of our catalog creation, we identify four sources of incompleteness: photometric detection, 2D-spectrum extraction with \texttt{grizli}, line detection with \texttt{grizli}, and visual inspection.

\subsection{Photometric detection completeness}

\begin{figure*}
    \centering
    \includegraphics[width=0.4\linewidth]{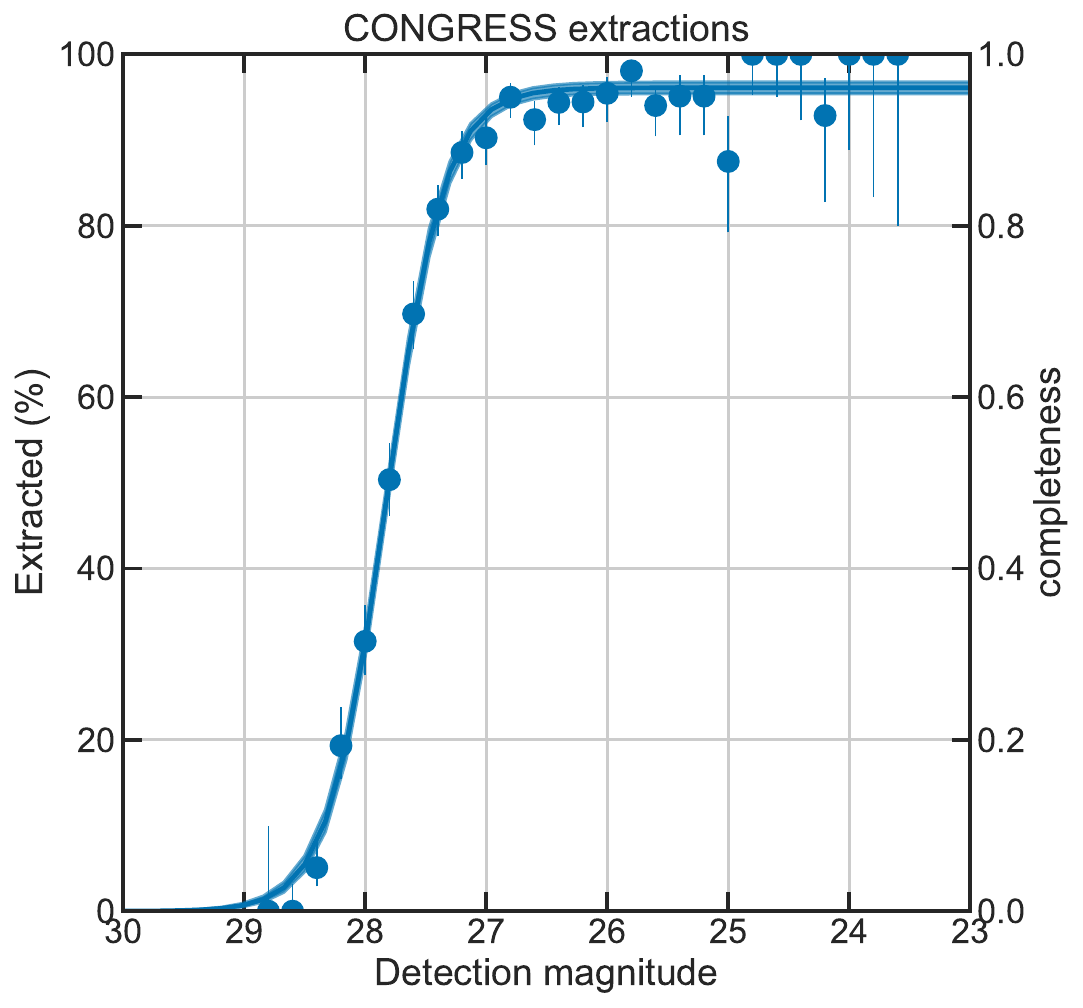}\includegraphics[width=0.4\linewidth]{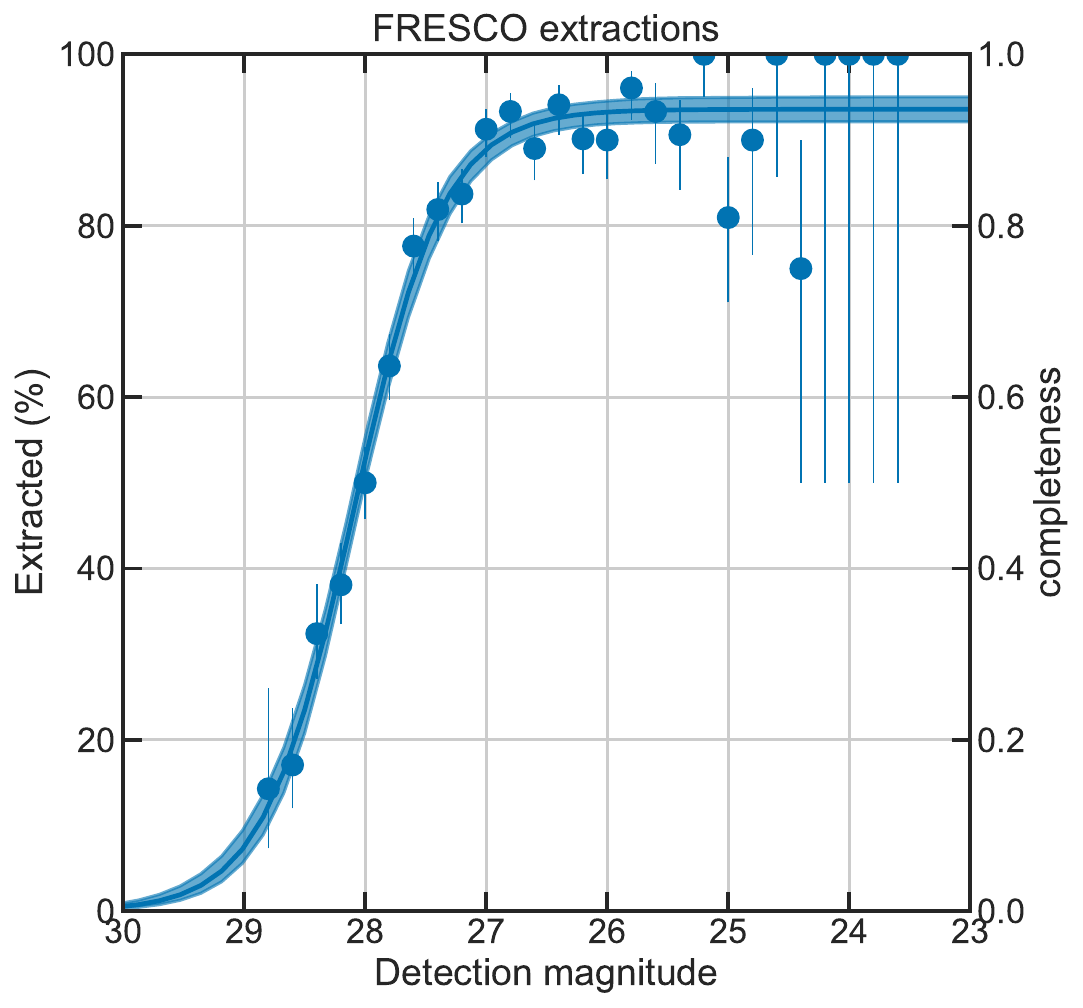}
    \caption{Percentage of sources that were extracted with \texttt{grizli}, in detection magnitude bins. The error bars represent the binomial confidence intervals. The extraction completeness is modeled as a sigmoid function, separately for CONGRESS (left) and FRESCO (right). The resulting sigmoid functions are shown in equations \ref{eq:comp_ext_1} and \ref{eq:comp_ext_2}. We reach maximum completeness at $\textrm{mag}\sim27$.}
    \label{Completeness_extraction}
     \end{figure*}

We assess the detection completeness of the photometric catalogs following the method described in \citet{Meyer+24},  which uses the GaLAxy survey Completeness AlgoRithm 2 (\texttt{GLACiAR2}; \citealt{Leethochawalit2022}). \texttt{GLACiAR2} performs injection-recovery simulations by creating synthetic galaxy profiles at a range of input magnitudes and morphological parameters, injecting them at random locations in the imaging mosaics and running \texttt{SourceExtractor} in the same way as for the real catalog to obtain the fraction of recovered sources as a function of the observed magnitude in the detection image, $\textrm{mag}_\textrm{det}$. As in \citet{Meyer+24}, the synthetic galaxies that we generated follow a log-normal distribution centered at $R_\textrm{eff}=0.8$ kpc, which was selected to fit the morphology of our photometric catalog in F444W and F210M. 
The completeness is well approximated by a sigmoid function. Thus, we fit a sigmoid function to the recovered sources as a function of $\textrm{mag}_\textrm{det}$, separately for each field, which yields the photometric detection completeness functions, $C_\textrm{det}$, already reported in \citet{Meyer+24} separately for GN and GS:

\begin{equation}
    C_\textrm{det;GN}= \frac{0.95}{1 + \textrm{exp}[1.61\times(\textrm{mag}_\textrm{det}-28.71)]};
    \label{eq:comp_det_1}
\end{equation}
\begin{equation}
    C_\textrm{det;GS}= \frac{0.93}{1 + \textrm{exp}[1.66\times(\textrm{mag}_\textrm{det}-28.55)]}.
    \label{eq:comp_det_2}
\end{equation}

\subsection{Extraction completeness}

During the line search process with \texttt{grizli} described in section \ref{sec:grism}, we only considered sources with photometric redshift uncertainties of less than $20\%$. This constitutes a source of incompleteness, as we did not extract sources with higher uncertainties, and some of these could still be H$\alpha$ emitters. As expected,  most of the sources with photometric redshifts in the target range that were not extracted were among the faintest targets in the photometric catalog. 
To account for this, we considered all the sources with an \texttt{eazy} photometric redshift whose redshift probability function overlaps significantly with the grism redshift sensitivity; i.e. we consider all the sources whose lower 16th percentile fulfills $z_{\textrm{phot},16}\geq3.72\,(4.86)$ for CONGRESS (FRESCO), and $z_{\textrm{phot},84}\leq5.15\,(6.69)$ for the upper 84th percentile. We binned these sources according to their detection magnitude and calculated the percentage of sources in each bin that were extracted. We then fitted this magnitude-dependent completeness again with a sigmoid function. The result is shown separately for CONGRESS and FRESCO in Fig. \ref{Completeness_extraction}, and the sigmoid functions that model this extraction completeness in each field are: 
\begin{equation}
    C_\textrm{ext;CONGRESS}= \frac{0.96}{1 + \textrm{exp}[4.17\times(\textrm{mag}_\textrm{det}-27.82)]};
    \label{eq:comp_ext_1}
\end{equation}
\begin{equation}
    C_\textrm{ext;FRESCO}= \frac{0.94}{1 + \textrm{exp}[2.70\times(\textrm{mag}_\textrm{det}-28.09)]}.
    \label{eq:comp_ext_2}
\end{equation}

Both of these functions cross the 50\% limit at about 28 mag.
 
\subsection{Line detection completeness}
Once the 2D-spectrum of each source is extracted, we use \texttt{grizli} to identify emission lines and extract their properties. Our final catalog only contains sources with an H$\alpha$ flux $\textrm{S/N}>5$, as we cannot reliably confirm that lower S/N detections are genuine H$\alpha$ emission lines rather than contamination. This leads to $1,388$ galaxies (662 in CONGRESS and 726 in FRESCO) that we are discarding, some of which might be genuine H$\alpha$ emitters.

To assess this line detection incompleteness, we simulate a catalog of $\sim 4,000$ mock H$\alpha$ emitters for each field and survey in a uniform grid of X and Y pixel position, emission line fluxes, and redshift. 
At each position, we extract the full grism spectrum (i.e., including possible contamination) and add synthetic emission lines with point source morphology. We generate synthetic emission lines for each mock source: one line of H$\alpha$ ($6\,564.59\AA$), a doublet of [NII] ($6\,549.84\AA$,  $6\,585.23\AA$), and a doublet of [SII] ($6\,718.32\AA$, $6\,732.71\AA$). The flux ratios for these five lines were set to $1:0.022:0.065:0.070:0.051$. These were generated for each source in a grid of wavelength positions and log(H$\alpha$) fluxes, covering the full range of observable wavelengths ($3.1-4.0\mu\textrm{m}$ for CONGRESS; $3.9-5.0\mu\textrm{m}$ for FRESCO) and observed H$\alpha$ fluxes ($8\times10^{-19}-8.5\times10^{-17}\textrm{erg}\,\textrm{s}^{-1}\textrm{cm}^{-2}$).

\begin{figure}
    \centering
    \includegraphics[width=\linewidth]{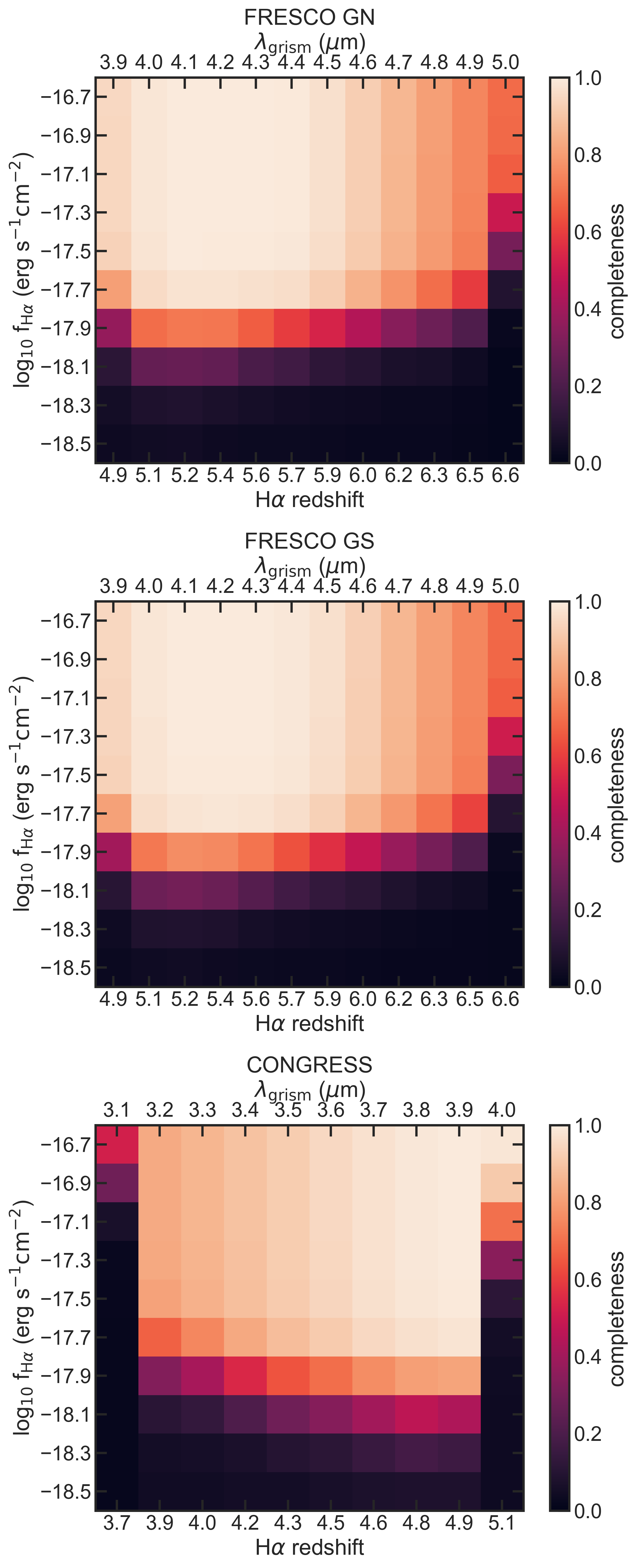}
\caption{Completeness of our line detection criteria as a function of grism wavelength (probing H$\alpha$ redshift) and H$\alpha$ flux, for each field and survey. Top: FRESCO GOODS-N. Middle: FRESCO GOODS-S. Bottom: CONGRESS (GOODS-N). These show an asymmetric wavelength dependence due to the sensitivity curves and cross the $50\%$ limit at $\log(\textrm{f}_{\textrm{H}\alpha}/$\fluxunit$)\sim-17.7$.}
    \label{Completeness_lines}
     \end{figure}

Finally, we used \texttt{grizli} to extract the mock H$\alpha$ emission lines for each source and each grid combination as for real sources. If the extracted H$\alpha$ line had a $\textrm{S/N}>5$, the completeness for that grid point was assigned 1; otherwise, the completeness was 0. We combined the resulting completeness grids of all the mock sources in each field and survey by using an average of all $\sim 4,000$ grids. This led to three separate completeness functions, each depending on two variables: the wavelength (i.e. redshift) and H$\alpha$ flux of the emission line. Although the completeness is also position-dependent, we marginalized over this dependence by placing sources uniformly across the FRESCO field and averaging the completeness array of all sources. 
These line-detection completeness functions are presented in Fig. \ref{Completeness_lines}. These show an asymmetric wavelength dependence due to the filter/grism throughput curve. In terms of H$\alpha$-flux level, the $50\%$ limit is generally crossed at $\log(\textrm{f}_{\textrm{H}\alpha}/$\fluxunit$)\sim-17.7$, which corresponds to the FRESCO $5\sigma$ sensitivity limit of $\sim2\times10^{-18}\textrm{erg}\,\textrm{s}^{-1}\,\textrm{cm}^{-2}$ as expected from the ETC \citep[see][]{Oesch23}. 

\subsection{Visual inspection completeness}
Due to possible contamination from other sources along the dispersion direction, grism spectra need to be visually inspected.
Experienced inspectors decide whether a line is likely to be a real H$\alpha$ emission line. However, some genuine emission lines might be missed during this process, due to a low S/N or the presence of contamination. To account for this visual inspection incompleteness, we selected a subsample of 320 mock H$\alpha$ emitters with $\textrm{S/N}>5$, separately for CONGRESS and FRESCO. For each survey, we selected mock sources at random pixel positions and redshifts in a grid of 
log(S/N). We did not cover the whole S/N range of our observations, as at the high-S/N end we found that emission lines are always properly identified during visual inspection and thus the bright sample is complete in this sense.

\begin{figure*}
    \centering
    \includegraphics[width=0.4\linewidth]{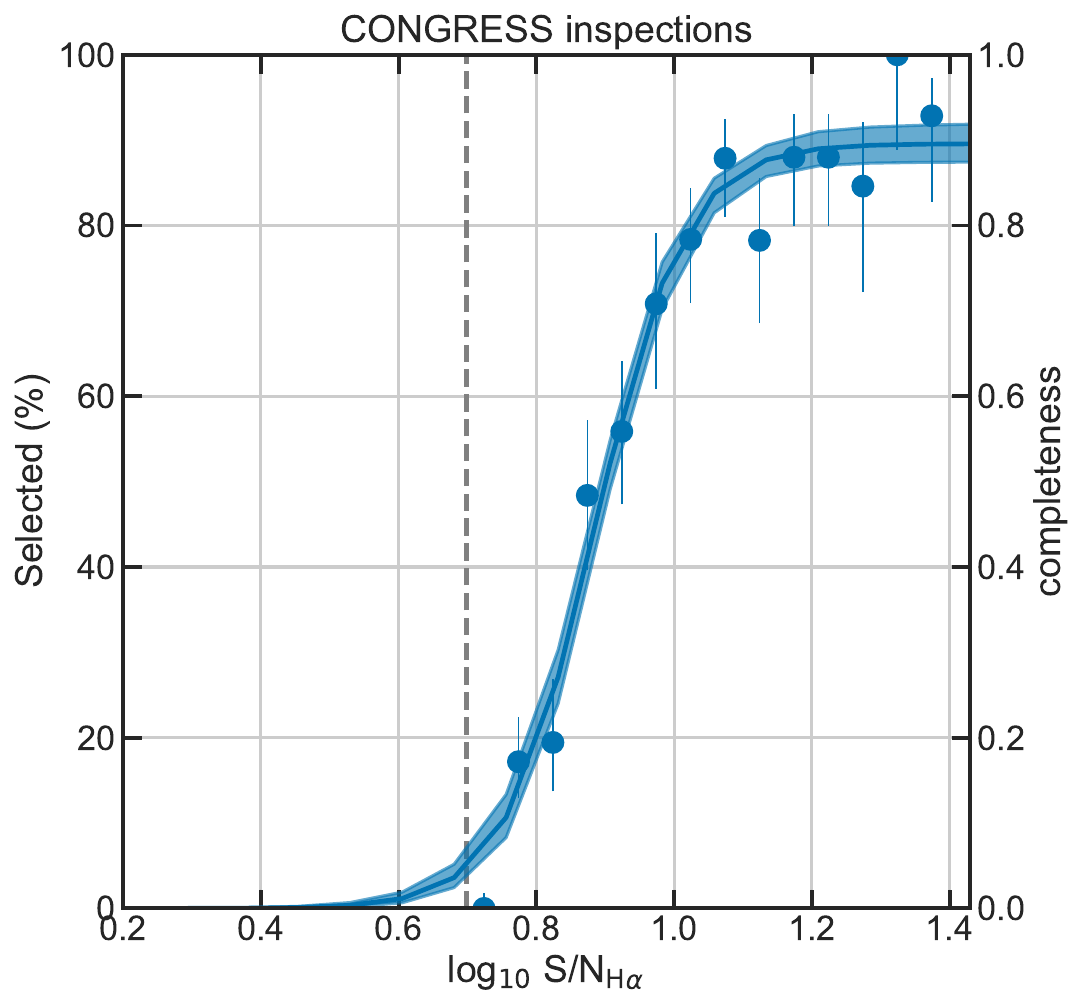}
    \includegraphics[width=0.4\linewidth]{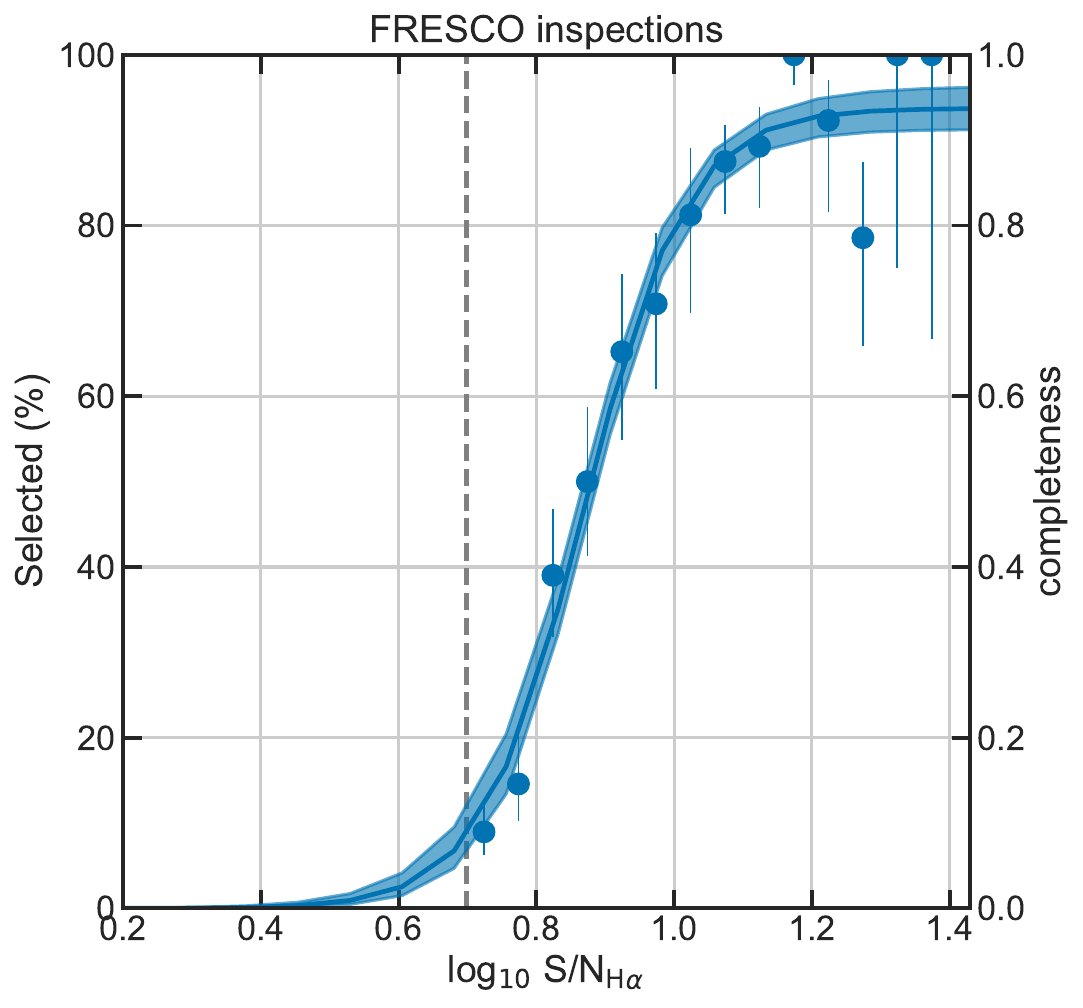}
    \caption{Percentage of mock sources that were selected through visual inspection as $q\geq1.75$, in log(H$\alpha$)-flux bins. The visual inspection completeness is modeled as a sigmoid function, separately for CONGRESS (left) and FRESCO (right). The vertical dashed line marks the S/N>5 limit. The resulting sigmoid functions are shown in equations \ref{eq:comp_ins_1} and \ref{eq:comp_ins_2}.}
    \label{fig:Completeness_inspection}
     \end{figure*}

We visually inspected these mock sources and gave them quality flags in the same manner as with the real data. Then, we computed the fraction of mock sources with $q\geq1.75$ in each flux bin, which correspond to the sources that would be selected into our sample if they were real. We fitted this 
S/N-dependent completeness with a sigmoid function. The result is shown separately for CONGRESS and FRESCO in  Fig. \ref{fig:Completeness_inspection}, and the sigmoid functions that model this visual inspection completeness in each field are:


\begin{equation}
    C_\textrm{ins;CONGRESS}= \frac{0.90}{1 + \textrm{exp}[15.46\times(\textrm{log}_\textrm{10}(\textrm{f}_{\textrm{H}\alpha})-0.89)]};
    \label{eq:comp_ins_1}
\end{equation}
\begin{equation}
    C_\textrm{ins;FRESCO}= \frac{0.94}{1 + \textrm{exp}[13.530.87\times(\textrm{log}_\textrm{10}(\textrm{f}_{\textrm{H}\alpha})-0.87)]},
    \label{eq:comp_ins_2}
\end{equation}

Both of these functions cross the $50\%$ limit at S/N$\sim7.5$.

\subsection{Total completeness}
\begin{figure}
    \centering
    \includegraphics[width=\linewidth]{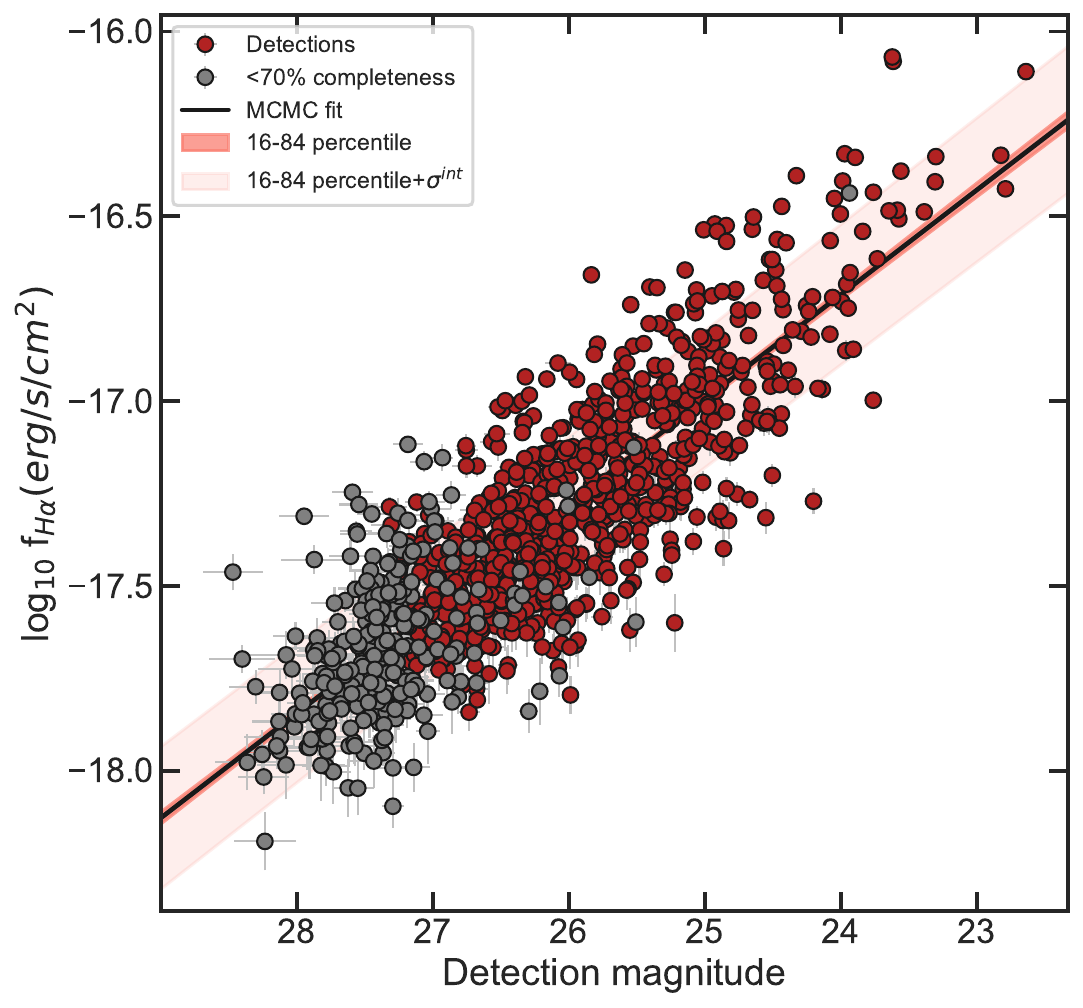}
\caption{The relation between our catalog's H$\alpha$ flux and the detection magnitude (stack of F444W and F210M). The solid, black line represents a Bayesian linear fit, while the shadowed area represents the intrinsic $1\sigma$ dispersion around the mean relation. Grey, filled circles are the sources with completeness below 65\%, while red circles correspond to galaxies with higher completeness. As can be seen, our catalog of H$\alpha$ emitters is starting to become incomplete at $\sim3\times10^{-18}$\fluxunit, where we are likely missing emitters that are not included in our parent catalog or are not extracted due to our requirements on photo-zs.}
    \label{fig:havsflux}
     \end{figure}

The total completeness is a multiplication
of the four completeness functions obtained in this section: photometric detection, extraction, line detection, and visual inspection. We compute the completeness for each source, which depends on the detection magnitude, H$\alpha$ flux, line wavelength (redshift), and H$\alpha$ S/N. The majority of the galaxies in our catalog have a total completeness above $65\%$, with only a quarter of them below this limit. In Fig. \ref{fig:havsflux}, we show the H$\alpha$ fluxes of galaxies as a function of the detection magnitude and we highlight the sources with an estimated completeness lower than $65\%$. 
As can be seen, completeness corrections start to become significant for sources below a detection magnitude of 27 mag and for fluxes fainter than  $3\times10^{-18}$ \fluxunit, corresponding to log $L_{\textrm{H}\alpha}/\textrm{erg}\;\textrm{s}^{-1}\lesssim41.75$; M$_\textrm{UV}\gtrsim-20$. Bins of luminosity with average completeness less than 65\% are not included in the following derivations of LFs.

\section{Results}
\label{sec:results}
\subsection{Redshift distribution and overdensities}

\begin{figure}
\centering
\includegraphics[width=\hsize]{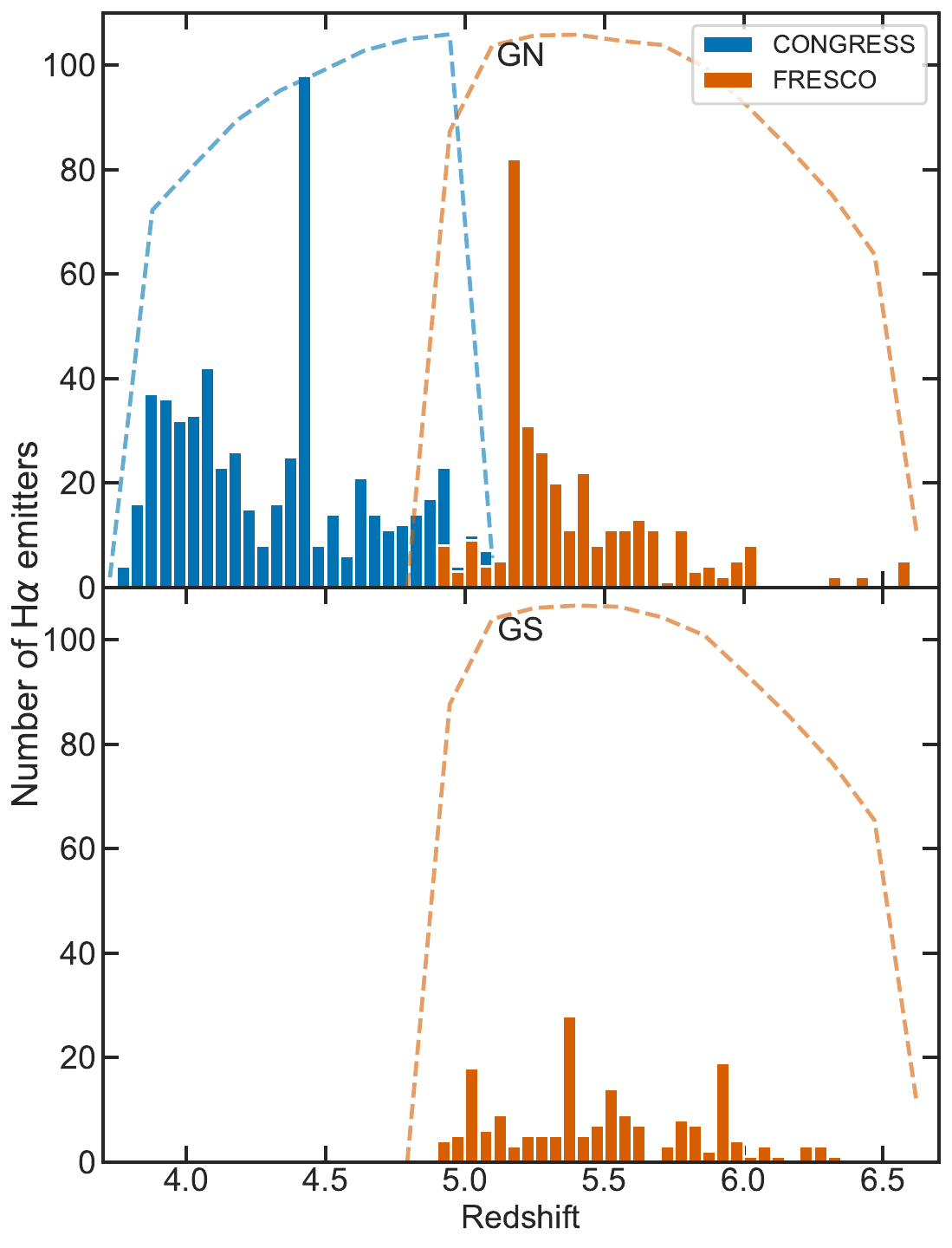}
\caption{Redshift distribution of the H$\alpha$ emitters found in CONGRESS (GOODS-N) and FRESCO (GOODS-N and GOODS-S). Dashed lines show the $\textrm{S/N}>5$ line detection completeness as a function of redshift for each survey and field. We find clear overdensities in GOODS-N at $z\sim4.4$ and $z\sim5.2$, and in GOODS-S at $z\sim5.4$ and $z\sim5.9$.}
\label{z_dist}
\end{figure}

We show the redshift distribution of our H$\alpha$ emitters in Fig. \ref{z_dist}, separately for each survey and field. From CONGRESS data, we find a significant overdensity in GOODS-N at $z\sim4.4$, with 98 out of all 554 H$\alpha$ emitters from the CONGRESS field lying at $z=4.43\pm$0.03. In the FRESCO fields, we find the same overdensities that were reported in  \citet{Helton+23}: at $z\sim5.4$ and $z\sim5.9$ in GOODS-S and a very large overdensity at $z\sim5.2$ in GOODS-N (see also \citealt{HerardDemanche24}).

As mentioned in section \ref{sec:selection}, among the H$\alpha$ emitters in our catalog, we found 13 sources consistent with AGN activity: five in CONGRESS and eight in FRESCO (reported in appendix \ref{sec:LRDs}). This constitutes a $1.09_{-0.28}^{+0.38}\%$ of such sources in our catalog. For CONGRESS, four out of the five AGN candidates belong to the overdensity at $z\sim4.4$, suggesting a possible connection between AGN and larger scale environment at the 3$\sigma$ level. We defer a more detailed discussion of this to a future paper.

\subsection{Observed H$\alpha$ luminosity function}
\label{sec:half}
We now compute the H$\alpha$ luminosity function at $3.8<z<6.6$. We split our full H$\alpha$ catalog into three redshift bins: one for all CONGRESS data, with $3.8<z<5.1$; and two for FRESCO data, splitting our redshift range in half: $4.9<z<5.7$, and $5.7<z<6.6$. We remove the broad-line H$\alpha$ emitters from the luminosity function calculation, as their H$\alpha$ flux does not only probe star formation (see Sect. \ref{sec:LRDs}). This resulted in 527 CONGRESS sources at $z\sim4.45$, 379 FRESCO sources at $z\sim5.3$, and 94 FRESCO sources in the last bin centered at $z\sim6.15$.

\begin{figure}
\centering
\includegraphics[width=\hsize]{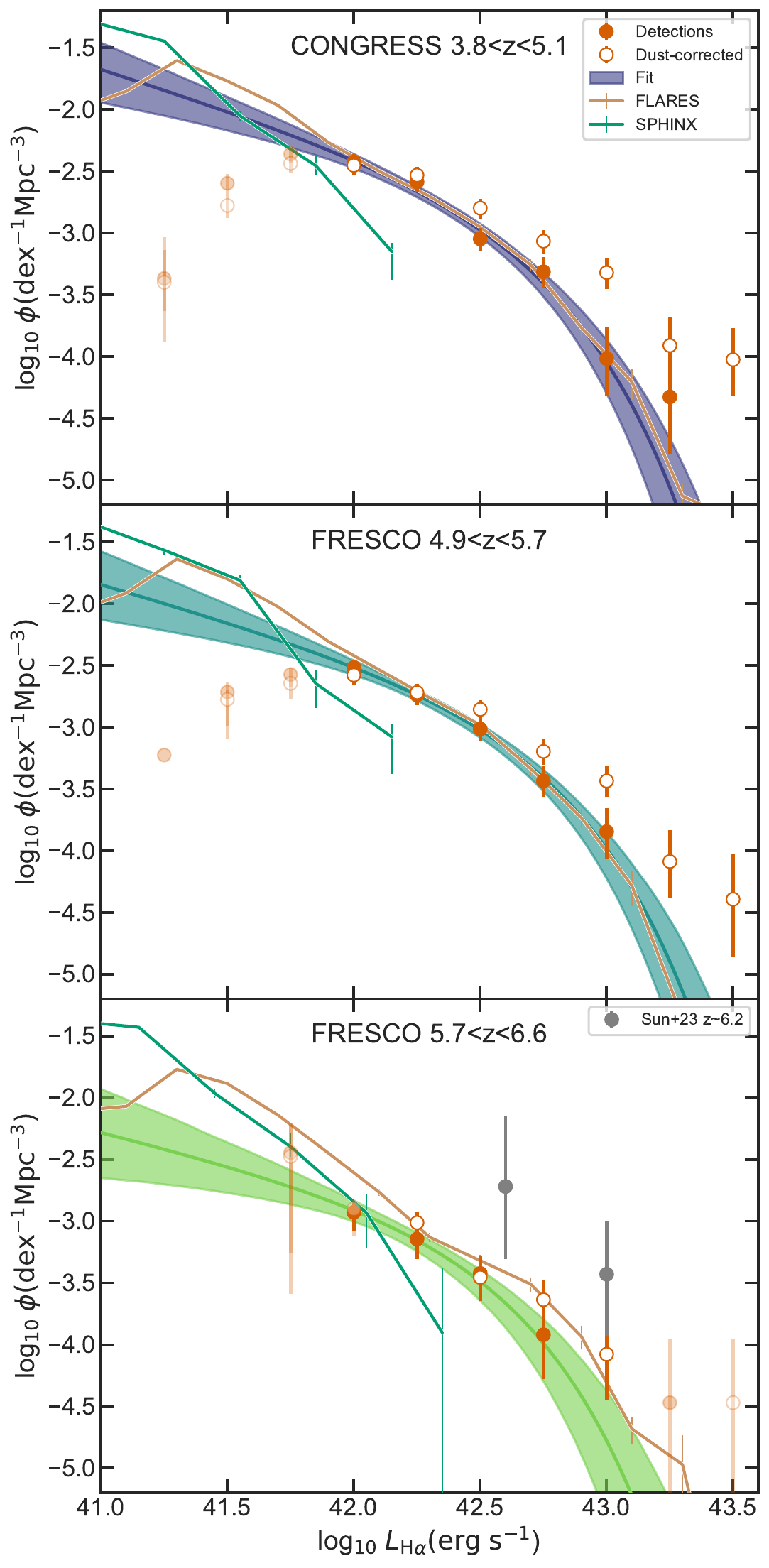}
\caption{The observed H$\alpha$ luminosity function at $z=3.8-6.6$, split in three redshift bins. Solid circles show the observed number densities, which are transparent when their average completeness is below 65\% (these values are not considered for the fit). Solid lines within shaded regions show the fitted Schechter functions and the 68\% confidence intervals. 
Also shown are the results of FLARES \citep{flares1,flares2} and SPHINX \citep{sphinx} 
at $z\sim4$, $z\sim5$, and $z\sim6$; and the results of \citet{Sun+23} at $z\sim6.2$.}
\label{HaLF}
\end{figure}

For each redshift bin, we derive the observed H$\alpha$ luminosity function following the $V/V_\textrm{max}$ method in \citet{Schmidt+68}. For each $\log L_{\textrm{H}\alpha}$ bin, we calculate the number density as 

\begin{equation}
    \phi\, d(\log L_{\textrm{H}\alpha}) =\sum_i\frac{V}{V_\textrm{max,i}C_{i}},
\label{eq:Vmax}
\end{equation}
where the summation $i$ runs over all galaxies in the given luminosity bin, $V$ is the full FRESCO survey volume over the redshift bin, $V_\textrm{max,i}$ is the maximum volume in which one could theoretically detect source $i$ in that $\log L_{\textrm{H}\alpha}$ bin at redshift $z_i$ with $\textrm{S/N}>5$, and $C_{i}$ is the completeness of the source $i$. The uncertainty of the number density includes Poisson noise, the completeness uncertainty, as well as cosmic variance following \citet{Trenti+08}, obtained with the cosmic variance calculator\footnote{Available at \url{https://www.ph.unimelb.edu.au/~mtrenti/cvc/CosmicVariance.html}}. The results for a $\log L_{\textrm{H}\alpha}$ bin width of 0.25 dex are shown in Tab. \ref{tab:HaLF} and Fig. \ref{HaLF}. We do not consider the number density values with an average completeness below 65\% for the fit, although we do show them in the figure as transparent points. We also include the dust-corrected H$\alpha$ luminosity function bins in Fig. \ref{HaLF}, which will be discussed in Sect. \ref{sec:sfrf}.

\begin{table}
    \centering
\renewcommand{\arraystretch}{1.3} 
    \begin{tabular}{cccc}
    \hline
     log $L_{\textrm{H}\alpha}$ & N & $<c>$ & $\phi_\textrm{comp}$\\
     $\textrm{erg}\;\textrm{s}^{-1}$&  &  & $10^{-3}\;\textrm{Mpc}^{-3}$ \\
     \hline
    \multicolumn{4}{c}{Congress ($z\sim4.45$)}\\
     \hline
42.00 & 149 & 0.70 & $3.75_{-0.59}^{+0.60}$ \\
42.25 & 108 & 0.73 & $2.59_{-0.43}^{+0.44}$ \\
42.50 & 38 & 0.75 & $0.90_{-0.19}^{+0.21}$ \\
42.75 & 20 & 0.73 & $0.48_{-0.13}^{+0.15}$ \\
43.00 & 4 & 0.73 & $0.096_{-0.048}^{+0.077}$ \\
43.25 & 2 & 0.75 & $0.047_{-0.031}^{+0.062}$ \\
    \hline
    \multicolumn{4}{c}{FRESCO ($z\sim5.3$)}\\
    \hline
42.00 & 135 & 0.75 & $3.04_{-0.47}^{+0.48}$ \\
42.25 & 87 & 0.80 & $1.82_{-0.30}^{+0.32}$ \\
42.50 & 47 & 0.82 & $0.97_{-0.19}^{+0.20}$ \\
42.75 & 18 & 0.82 & $0.366_{-0.098}^{+0.118}$ \\
43.00 & 7 & 0.83 & $0.142_{-0.056}^{+0.079}$ \\
    \hline
    \multicolumn{4}{c}{FRESCO ($z\sim6.15$)}\\
    \hline
42.00 & 36 & 0.68 & $1.18_{-0.34}^{+0.23}$ \\
42.25 & 23 & 0.69 & $0.72_{-0.22}^{+0.17}$ \\
42.50 & 13 & 0.70 & $0.38_{-0.12}^{+0.15}$ \\
42.75 & 4 & 0.75 & $0.120_{-0.067}^{+0.075}$ \\
    \hline
    \end{tabular}
    \caption{H$\alpha$ luminosity function at $z\sim4.45$, $z\sim5.3$, and $z\sim6.15$: number of emitters, average completeness, and completeness-corrected number densities for each luminosity bin. The completeness is a multiplication of the detection, extraction, line detection, and visual inspection completeness computed for each source (see section \ref{sec:completeness}).}
    \label{tab:HaLF}
\end{table}

We fit a Schechter function to the observed H$\alpha$ luminosity functions:
\begin{equation}
    \phi=\frac{dn}{d\log{L}}=\ln{(10)}\,\phi^*\,10^{(\log{L}-\log{L^*})\,(\alpha +1)}\textrm{exp}\left[10^{(\log L -\log L^*)}\right],
\end{equation}
parametrized by $\phi^*$ as the characteristic number density, $L^*$ as the characteristic luminosity, and $\alpha$ as the faint-end slope. To estimate these parameters, we employed a Bayesian inference approach using Markov Chain Monte Carlo (MCMC) sampling with the \texttt{pymc} package. We set weakly informative priors for the parameters chosen as normal distributions centered on the values found by \citet{Bollo23} at $z\sim4.4$. We started the sampling process by finding the parameter values that maximize the posterior distribution. Subsequently, we performed MCMC sampling with $5\,000$ iterations across four chains to obtain posterior distributions for the parameters. The posteriors were used to derive the median values and 16th-84th percentiles for each parameter, providing a comprehensive statistical characterization of the fitted Schechter function. The results of these fits are listed in Tab. \ref{tab:HaLF_fit} and also shown in Fig. \ref{HaLF}. 

\begin{table}
    \centering
\renewcommand{\arraystretch}{1.3} 
    \begin{tabular}{cccc}
    \hline
     $z$ & log $L^*_{\textrm{H}\alpha}$ & $\phi^*_{\textrm{H}\alpha}$ & $\alpha_{\textrm{H}\alpha}$\\
     & $\textrm{erg}\;\textrm{s}^{-1}$& $10^{-3}\;\textrm{Mpc}^{-3}$ &\\
     \hline
4.45&$42.60^{+0.17}_{-0.16}$ & $0.89^{+0.69}_{-0.43}$ & $-1.64^{+0.27}_{-0.21}$ \\
5.30&$42.65^{+0.18}_{-0.17}$ & $0.69^{+0.55}_{-0.35}$ & $-1.58^{+0.28}_{-0.25}$ \\
6.15&$42.46^{+0.24}_{-0.20}$ & $0.45^{+0.40}_{-0.25}$ & $-1.49^{+0.36}_{-0.33}$ \\
    \hline
    \end{tabular}
    \caption{Schechter parameters of the observed H$\alpha$ luminosity functions at $z\sim4.45$, $z\sim5.3$, and $z\sim6.15$.}
    \label{tab:HaLF_fit}
\end{table}

\subsection{Simulated H$\alpha$ luminosity function}
In order to compare the observed H$\alpha$ luminosity functions with theoretical predictions, we added several results from simulations to Fig. \ref{HaLF}. In particular, we use the results of the First Light And Reionization Epoch Simulations (FLARES) \citep{flares1,flares2} and SPHINX$^{20}$ \citep{Rosdahl_22}, specifically the SPHINX Public Data Release v1 (SPDRv1) catalog of mock observations described by \cite{sphinx}.

While there are significant differences between the predicted LFs from these three simulations, we find decent agreement overall with our observational estimates. In particular, we find very good agreement with the predictions from FLARES, especially at the bright end. Nevertheless, in the highest redshift bin, the FLARES LF shifts towards the higher end of our fit. 
Interestingly, all three simulations consistently predict a higher number density of sources below our detection limit (at $\log L_{\mathrm{H}\alpha}/\mathrm{erg}\,\mathrm{s}^{-1} \sim 41.5$) than the extrapolation of our derived LFs. This might indicate that our faint end slopes are somewhat too shallow and deeper observations are needed to test this. However, we note that the overall shape of our observed LFs, including the faint-end slopes, are in very good agreement with the predictions from the SC SAM model, except that they are offset by up to $\sim0.3-0.5$dex in normalization. 
On the other hand, we find that SPHINX underpredicts the LF, where it overlaps with our sample, which only occurs over a very small luminosity range due to the smaller volume of SPHINX.
The different results between the simulations probably stem from different approaches to the modeling of line emission, mostly separating SPHINX from FLARES and SC SAM. SPHINX can model the interstellar medium and self-consistently compute nebular emission; however, this is done at the expense of box size, which means that bright galaxies such as those observed by FRESCO and CONGRESS are rare, hence the large uncertainties in the bright end of the LF. In contrast, FLARES and SC SAM model emission lines through the use of Cloudy \citep{ferland98, Chetzikos23}. Similar work in comparing high redshift line emission LFs with simulations was done by \cite{Meyer+24}, who also found good agreement with FLARES and SC SAM, and that SPHINX underpredicts their values. These comparisons show that the summary statistics of emission line measurements like the LFs are a powerful new avenue to test early galaxy formation models. 

\subsection{UV luminosity function}

\begin{figure}
\centering
\includegraphics[width=\hsize]{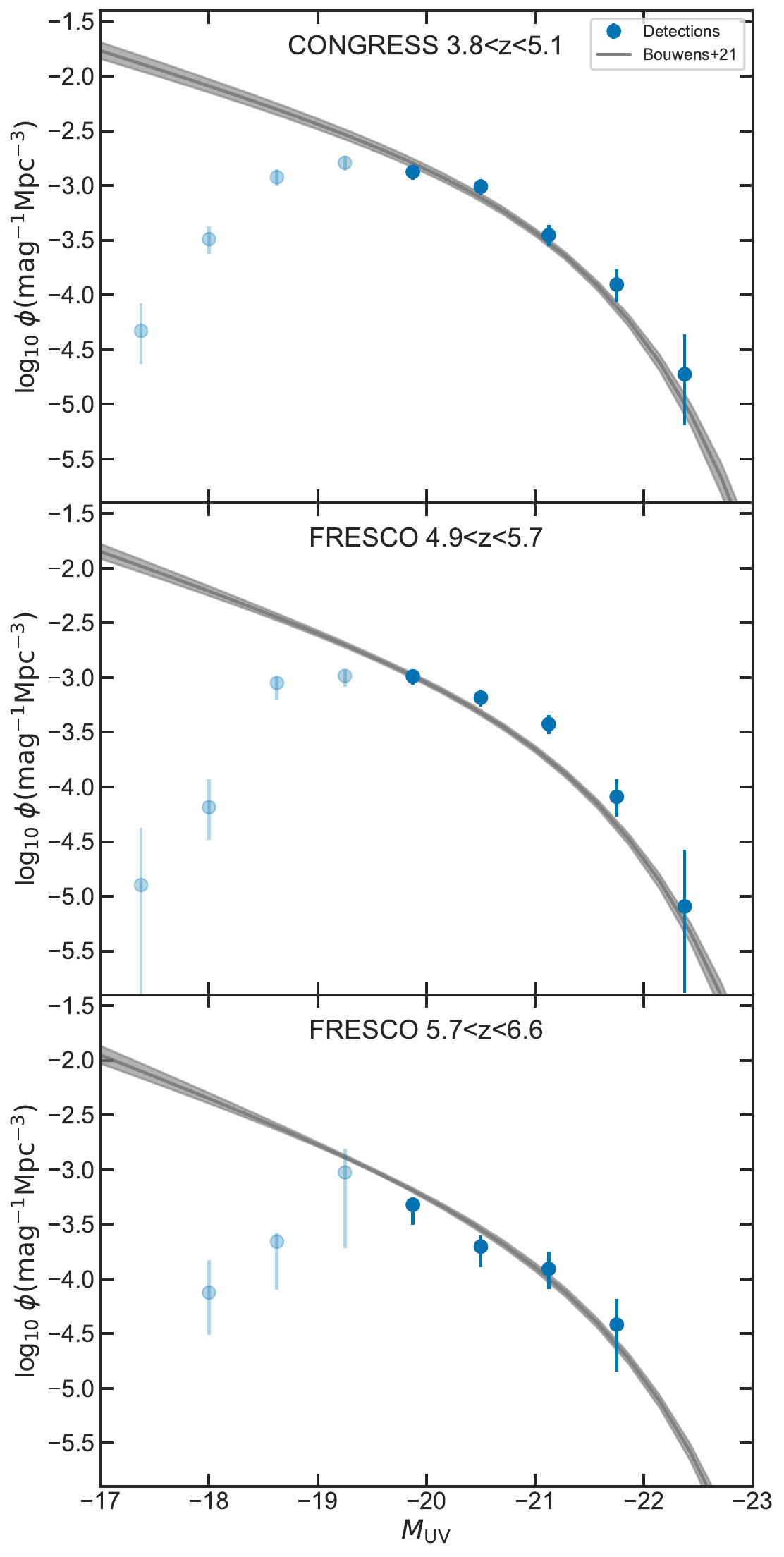}
\caption{The GOODS fields UV luminosity function at $z=3.8-6.6$, split into three different redshift bins. Solid circles show the observed number densities, which are transparent when their average completeness is below 65\%. Grey lines and shaded regions are the UV luminosity functions with 68\% confidence intervals obtained by \citet{Bouwens21} for redshifts $4.45$, $5.3$, and $6.15$, respectively.}
\label{UVLF}
\end{figure}

Our new catalog of H$\alpha$ emitters in the GOODS fields allows us to revisit the UV LF measurements that have been performed over the last decade with photometrically selected Lyman-break galaxies, but now for the first time with spectroscopically confirmed sources. Using the same methods as in section \ref{sec:half}, we divided our sources into the same three redshift bins and used a $M_\textrm{UV}$ bin size of 0.625 mag, akin to the $\log L_{\textrm{H}\alpha}$ bin separation. The results are shown in Fig. \ref{UVLF}, plotted together with the UV LF Schechter fits from \citet{Bouwens21}. We find that, for $M_\textrm{UV}$ bins where the average completeness is above 65\%, our results based on spectroscopic redshifts are an excellent match, within uncertainty, to those obtained by \citet{Bouwens21} using photometrically-selected objects. This provides further confidence in the accuracy of previous UV LF measurements at these epochs.

\subsection{Star formation rate functions}
\label{sec:sfrf}

\begin{figure}
\centering
\includegraphics[width=\hsize]{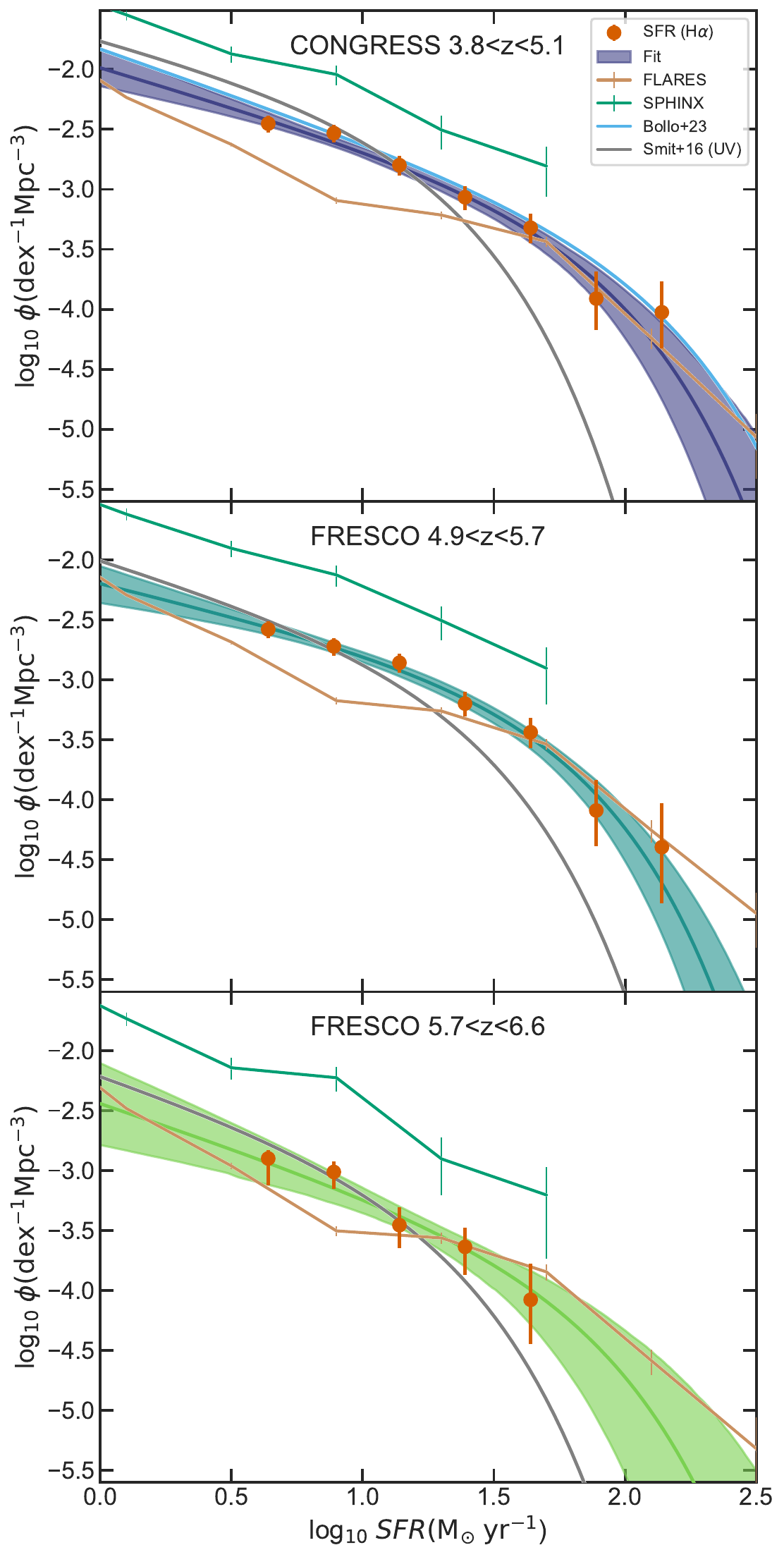}
\caption{The star formation rate functions at $z=3.8-6.6$, split in three redshift bins. Solid circles show the number densities derived from dust-corrected H$\alpha$ luminosities. Solid lines within shaded regions show the fitted Schechter functions and the 68\% confidence intervals for H$\alpha$ values. Also shown are the simulations results of FLARES \citep{flares1,flares2} and SPHINX \citep{sphinx} at $z\sim4$, $z\sim5$, and $z\sim6$; and the literature H$\alpha$-derived results from \citet{Bollo23} at redshifts $3.86<z<4.94$ and UV-derived from \citet{Smit+16} at $z\sim3.8$, $4.9$, and $5.9$ (scaled to match our H$\alpha$ conversion factor)}
\label{SFRF}
\end{figure}

The UV and H$\alpha$ LFs are based on observed quantities. However, what we really would like to constrain is the build-up of star formation in early galaxies at $z>3$. We can now do this based on the observed H$\alpha$ emission lines, after a correction for dust extinction. 
To convert the H$\alpha$ luminosities to SFRs, 
we used the relation in \citet{Kennicutt+98}, scaled by a factor of 1.8 to consider a \citet{Chabrier+03} Initial Mass Function:
\begin{equation}
    \textrm{SFR}_{\textrm{H}\alpha}\;(\textrm{M}_\odot\; \textrm{yr}^{-1}) =  10^{-41.36}\;L_{\textrm{H}\alpha,\textrm{int}} \;(\textrm{erg}\;\textrm{s}^{-1}).
    \label{eq:ha-sfr}
\end{equation}

We computed the dust-corrected H$\alpha$ luminosity using intrinsic H$\alpha$ luminosities, obtained with:
\begin{equation}    L_{\textrm{H}\alpha\textrm{,int}}=L_{\textrm{H}\alpha\textrm{,obs}}\times10^{0.4\, A_{\textrm{H}\alpha}},
\end{equation}
where $A_{\textrm{H}\alpha}$ is the dust attenuation at $6\,564.6$ $\AA$ derived in section \ref{sec:sed}. We then computed the SFR functions in the same way as the H$\alpha$ LFs (see Eq. \ref{eq:Vmax}). This led to the H$\alpha$-derived star formation rate functions reported in Tab. \ref{tab:SFRF} and shown in Fig. \ref{SFRF}. We then fitted Schechter functions to these values, using the same MCMC techniques as with the H$\alpha$ luminosity function. The results of these fits are shown in Tab. \ref{tab:SFRF_fit}.

\begin{table}
    \centering
\renewcommand{\arraystretch}{1.3} 
    \begin{tabular}{ccc}
    \hline
     log SFR & N & $\phi_\textrm{SFR}$\\
     $\textrm{M}_\odot\;\textrm{yr}^{-1}$&   & $10^{-3}\;\textrm{Mpc}^{-3}$\\
     \hline
    \multicolumn{3}{c}{Congress ($z\sim4.45$)}\\
     \hline
0.64 & 137 & $3.53_{-0.56}^{+0.58}$ \\
0.89 & 121 & $2.94_{-0.48}^{+0.49}$ \\
1.14 & 66 & $1.59_{-0.29}^{+0.31}$ \\
1.39 & 36 & $0.86_{-0.18}^{+0.20}$ \\
1.64 & 20 & $0.48_{-0.12}^{+0.15}$ \\
1.89 & 5 & $0.123_{-0.056}^{+0.085}$ \\
2.14 & 4 & $0.094_{-0.047}^{+0.076}$ \\
    \hline
    \multicolumn{3}{c}{FRESCO ($z\sim5.3$)}\\
    \hline
0.64 & 112 & $2.64_{-0.42}^{+0.43}$ \\
0.89 & 88 & $1.91_{-0.32}^{+0.33}$ \\
1.14 & 67 & $1.39_{-0.25}^{+0.26}$ \\
1.39 & 31 & $0.63_{-0.14}^{+0.16}$ \\
1.64 & 18 & $0.366_{-0.098}^{+0.118}$ \\
1.89 & 4 & $0.082_{-0.040}^{+0.065}$ \\
2.14 & 2 & $0.040_{-0.027}^{+0.053}$ \\
    \hline
    \multicolumn{3}{c}{FRESCO ($z\sim6.15$)}\\
    \hline
0.89 & 32 & $0.97_{-0.27}^{+0.22}$ \\
1.14 & 11 & $0.35_{-0.12}^{+0.14}$ \\
1.39 & 8 & $0.231_{-0.097}^{+0.102}$ \\
1.64 & 3 & $0.084_{-0.048}^{+0.083}$ \\
    \hline
    \end{tabular}
    \caption{Star formation rate function from H$\alpha$ emitters at $z\sim4.45$, $z\sim5.3$, and $z\sim6.15$: number of emitters and number density for each star-formation bin.}
    \label{tab:SFRF}
\end{table}

\begin{table}
    \centering
\renewcommand{\arraystretch}{1.3} 
    \begin{tabular}{ccccc}
    \hline
     $z$ & log SFR$^*$ & $\phi^*_{\textrm{SFR}}$ & $\alpha_{\textrm{SFR}}$ & $\rho_{\textrm{SFR;H}\alpha}$\\
     & $\textrm{M}_\odot\;\textrm{yr}^{-1}$& $10^{-3}\textrm{Mpc}^{-3}$ & &$\textrm{M}_\odot\;\textrm{yr}^{-1}\textrm{Mpc}^{-3}$\\
     \hline
4.45&$2.03^{+0.32}_{-0.21}$ & $0.33^{+0.34}_{-0.22}$ & $-1.64^{+0.20}_{-0.19}$ & $0.041^{+0.005}_{-0.006}$\\
5.30&$1.89^{+0.19}_{-0.17}$ & $0.40^{+0.30}_{-0.19}$ & $-1.52^{+0.20}_{-0.18}$ & $0.029^{+0.003}_{-0.004}$\\
6.15&$1.98^{+0.38}_{-0.34}$ & $0.09^{+0.16}_{-0.06}$ & $-1.73^{+0.30}_{-0.26}$ & $0.012^{+0.003}_{-0.003}$\\
    \hline
    \end{tabular}
    \caption{Schechter parameters of the H$\alpha$-derived star formation rate functions at $z\sim4.45$, 
    $z\sim5.3$, and $z\sim6.15$. The last column is the SFRD, which results from integrating these Schechter functions.}
    \label{tab:SFRF_fit}
\end{table}
 
At all redshifts, we find characteristic SFRs of $\sim100$ M$_\odot$/yr. This is significantly higher than the characteristic SFRs previously derived by \citet{Smit+16} for the SFR functions at these redshifts based on H$\alpha$ lines inferred from Spitzer/IRAC broad-band photometry. We also show these estimates in Fig. \ref{SFRF}. As can be seen, we find a significantly higher number density of galaxies with high SFRs ($\log\mathrm{SFR}/\mathrm{M}_\odot\mathrm{yr}^{-1}>2$) compared to the previous estimates from \citet{Smit+16}. This difference is likely not the result of wrongly estimated emission line fluxes from the photometric data; it is more likely a consequence of different techniques: while we derive the SFR function directly from individual dust-corrected H$\alpha$ luminosities, \citet{Smit+16} used their H$\alpha$-based SFRs to derive an average dust correction to convert the UV LFs to SFR functions. We suspect that this procedure likely underestimates the number of high-SFR sources that are present due to the significant amount of scatter in the $L_{UV}-L_{\mathrm{H}\alpha}$ relation. Additionally, the UV-based estimates probe star-formation on longer timescales than H$\alpha$, which leads to smaller dispersion. 
This is supported by the fact that our estimates are in very good agreement with the ones from \citet{Bollo23} who also derived SFR functions at $z\sim4.5$ based on H$\alpha$ lines inferred from Spitzer/IRAC broad-band photometry, but who used a similar procedure as we adopt here to derive the SFR functions for individual sources.

\begin{figure*}
\centering
\includegraphics[width=14cm]{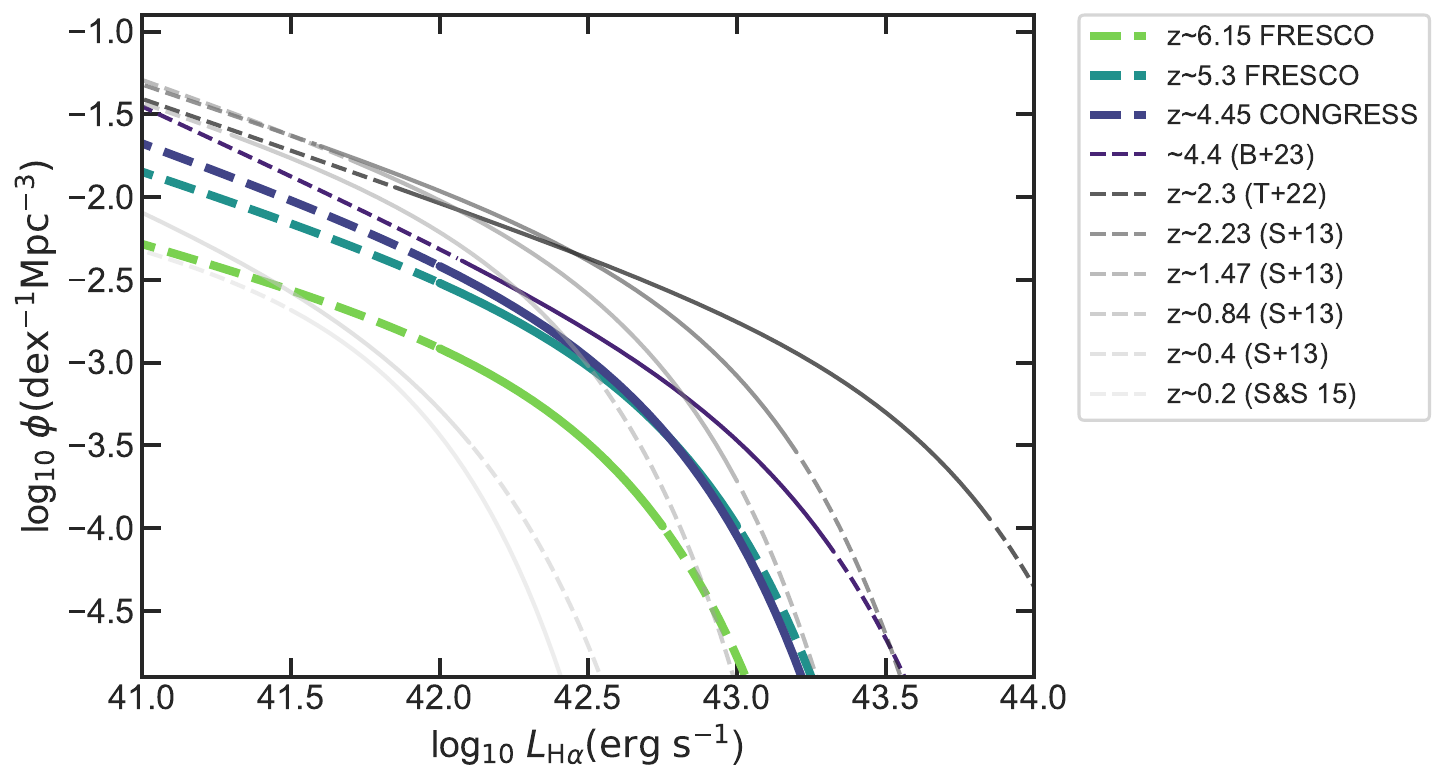}
\caption{The evolution of the H$\alpha$ LFs across cosmic history. We compare our best-fit H$\alpha$ LFs at $z\sim4.45$, $z\sim5.3$, and $z\sim6.15$, to previous results: \citet{Stroe&Sobral15} at $z\sim0.2$; \citet{Sobral+13} at $z\sim$0.4, 0.84, 1,47, and 2.23; \citet{Terao+22} at $z\sim2.3$; and \citet{Bollo23} at $z\sim4.4$. Solid lines represent the regions that were probed observationally in each study. For those studies that only reported the dust-corrected luminosity functions, we inferred the uncorrected functions by reversing the dust
correction applied for a fair comparison. We observe a trend of number densities increasing with decreasing redshift up to cosmic noon, to then reverse this trend and decrease until the present universe.}
\label{HaLF_comp}
\end{figure*}

We can further compare the derived SFR functions with simulations. In particular, we have computed the SFR functions directly from the SPHINX and FLARES data. Once again, we find relatively good agreement with FLARES over most of the SFR range. Albeit, the FLARES SFR functions show a dip at $\mathrm{SFR}=10\,\mathrm{M}_\odot\mathrm{yr}^{-1}$ at all redshifts, where their number densities are lower than what we observe by up to 0.5 dex. 
For the SPHINX simulation, the situation is different. While it predicts fainter H$\alpha$ lines than we observe, the SFR functions are higher than what we find by more than 0.5 dex. This overestimation of the SFR function is similar to the one seen in the stellar mass function \citep[see][]{Weibel2024}. Taken together, this likely indicates that a large amount of dust is present in the simulated galaxies, which obscures the observed quantities such as the H$\alpha$ \citep[or the UV fluxes; see also][]{sphinx}, leading to an underpredicted H$\alpha$ LF even though the intrinsic number density of galaxies at a given SFR is higher than observed. 

\subsection{Star formation rate densities}
Equipped with the SFR functions derived in the previous section, we can now compute the SFRDs from our H$\alpha$ measurements. We do this by integrating the SFR function, separately for each of the three redshift bins. For consistency with previous work, we set the lower integration limit to $\textrm{SFR}_l=0.27\,M_\odot\,\textrm{yr}^{-1}$, corresponding to a UV luminosity limit of $M_{\textrm{UV},l}=-17$. This limit was also used in \citet{Bouwens+15}, that we use to compare the UV and H$\alpha$-based SFRD estimates. As the integrands are Schechter functions, we can solve these integrals analytically using: 
\begin{equation}
\begin{split}    \rho_{\textrm{SFR;H}\alpha}=\int_{\log \textrm{SFR}_l}^\infty \phi(\log \textrm{SFR})\, \textrm{SFR}\,\textrm{d}(\log \textrm{SFR})=\\
    \phi_\textrm{SFR}^*\, \textrm{SFR}^*\,\Gamma\left(\alpha+2,\textrm{SFR}_l/\textrm{SFR}^*\right),
\end{split}   
\end{equation}
where $\Gamma$ is the incomplete gamma function:
\begin{equation}
\Gamma(a,x)=\int_x^\infty t^{a-1}\,e^{-t}\,dt.
\end{equation}
For each redshift bin, we integrated each of the $5\,000$ posterior Schechter function parameters obtained during the MCMC sampling and used the median of these results as the SFRD value, and the 16th-84th percentiles as the uncertainties.

This yields the SFRDs reported in Tab. \ref{tab:SFRF_fit} for each redshift bin: at $z\sim4.45$ we find $0.042\,M_\odot\,\mathrm{yr}^{-1}\,\mathrm{Mpc}^{-3}$ and this declines by a factor of 3 towards $z\sim6.15$. As will be discussed in the following section, this is in good agreement with dust-corrected, UV-based measurements \citep[e.g.,][]{Bouwens+15}.

\section{Discussion}
\label{sec:discussion}
\subsection{H$\alpha$ luminosity function evolution}
\begin{figure}
\centering
\includegraphics[width=\hsize]{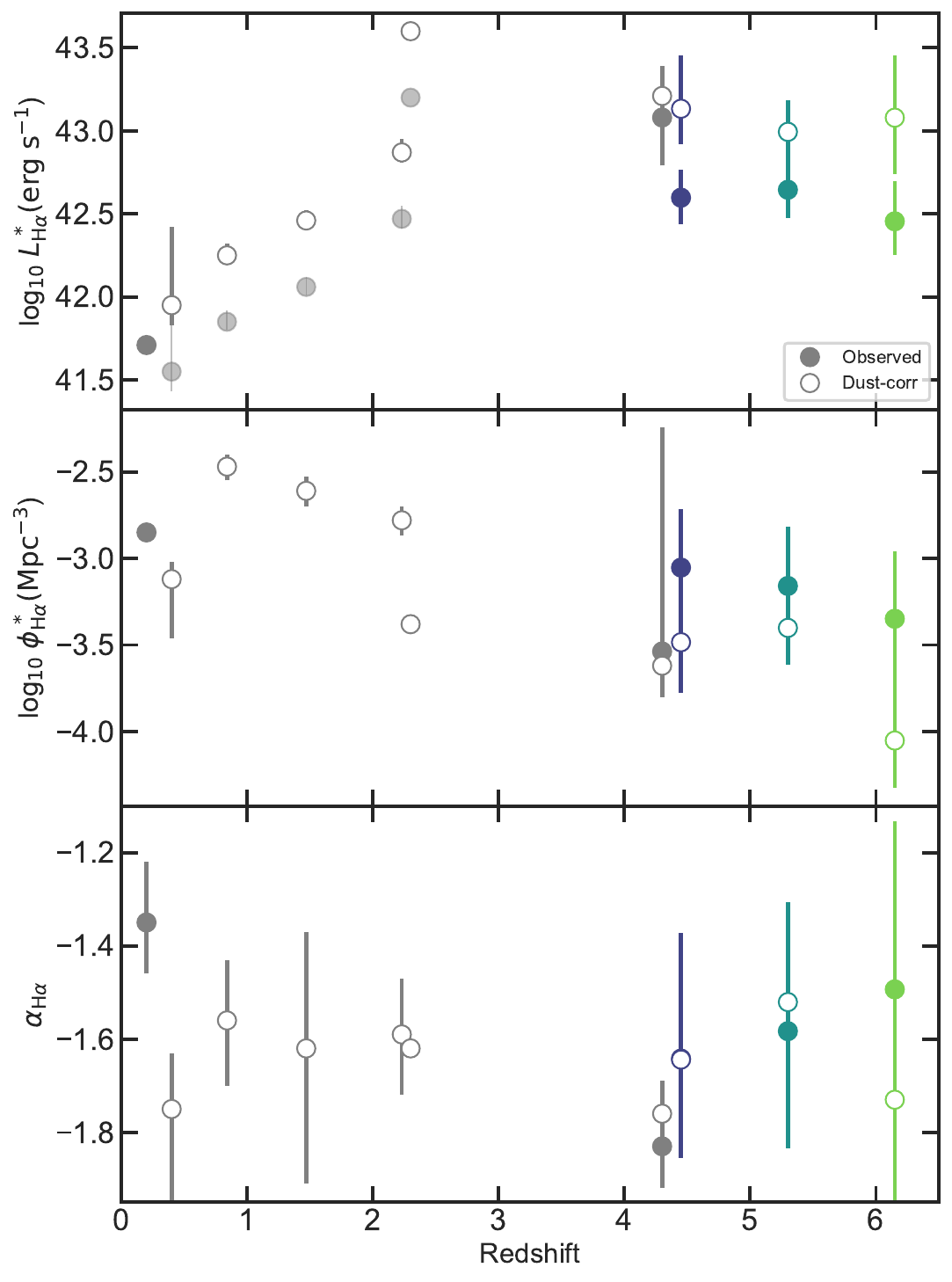}
\caption{Cosmic evolution of the H$\alpha$ LF Schechter parameters: our best-fit values compared to previous results at lower redshifts: \citet{Stroe&Sobral15} at $z\sim0.2$; \citet{Sobral+13} at $z\sim$0.4, 0.84, 1,47, and 2.23; \citet{Terao+22} at $z\sim2.3$; and \citet{Bollo23} at $z\sim4.4$. Filled circles represent the parameters of observed H$\alpha$ LFs, while open circles are for dust-corrected fits. For those studies that only reported the Schechter parameters of dust-corrected measurements, we inferred the $\log L^*$ parameter by reversing the dust correction applied, and show those values as fainter solid circles.}
\label{Param_comp}
\end{figure}

Fig. \ref{HaLF_comp} shows the redshift evolution of the H$\alpha$ LF across cosmic history from $z\sim 0.2$ up to $z\sim6.1$ based on literature estimates as well as our new results.
We observe an evolution of these luminosity functions, where $z\sim6$ galaxies are $\sim1$ dex less numerous than those at cosmic noon. This coincides with the epoch of galaxy build-up, hence it is expected that galaxy numbers increase with decreasing redshift.
We also observe that the bright end of the H$\alpha$ luminosity function is low ($\log (L_{\textrm{H}\alpha}/\textrm{erg}\;\textrm{s}^{-1})\sim42.5$) at local redshifts ($0.2<z<0.4$), and it increases with redshift up to a maximum of $\log (L_{\textrm{H}\alpha}/\textrm{erg}\;\textrm{s}^{-1})\sim44$ at $z\sim2-3$, to then decrease back to $\log (L_{\textrm{H}\alpha}/\textrm{erg}\;\textrm{s}^{-1})\sim42.5$. Given that H$\alpha$ is a tracer of galaxy star formation rate, this behavior agrees with the evolution shown by \citet{MD14}, where there is a peak of star formation at $z\sim2-3$.
This is also noticeable in Fig. \ref{Param_comp}, where we compare our best-fit Schechter parameters to those of previous studies at $z\sim0.2-4.4$. The characteristic luminosity $\log L^*$ follows the increasing trend  with redshift up to $z\sim3$, to then remain fairly constant for all $4<z<6$. Even though we note that there are degeneracies in Schechter function fits, the trends in Fig. \ref{Param_comp} indicate that the key driver for the evolution of the H$\alpha$ LFs at $z<3$ is the characteristic luminosity, which then turns to an evolution in density at $z>4$, i.e., the normalization parameter $\log\phi^*$ shows a slight decrease with increasing redshift, in agreement with previous results \citep{Bollo23}.

At the same time, we do not find any significant evolution of the faint-end slope, $\alpha$, with redshift, unlike for the UV luminosity functions reported by \citet{Bouwens21}, where this parameter tends to flatten with decreasing redshift, being as steep as $\alpha\sim-2$ at $z\sim7$. However, the uncertainty in our faint-end slope estimates is still very large. Deeper grism exposures will be required to constrain this better in the future, such as the ones from the ALT survey (Naidu, Matthee, et al., 2024 in prep.).

\subsection{Star formation rate density evolution}
\begin{figure*}
\centering
\includegraphics[width=12cm]{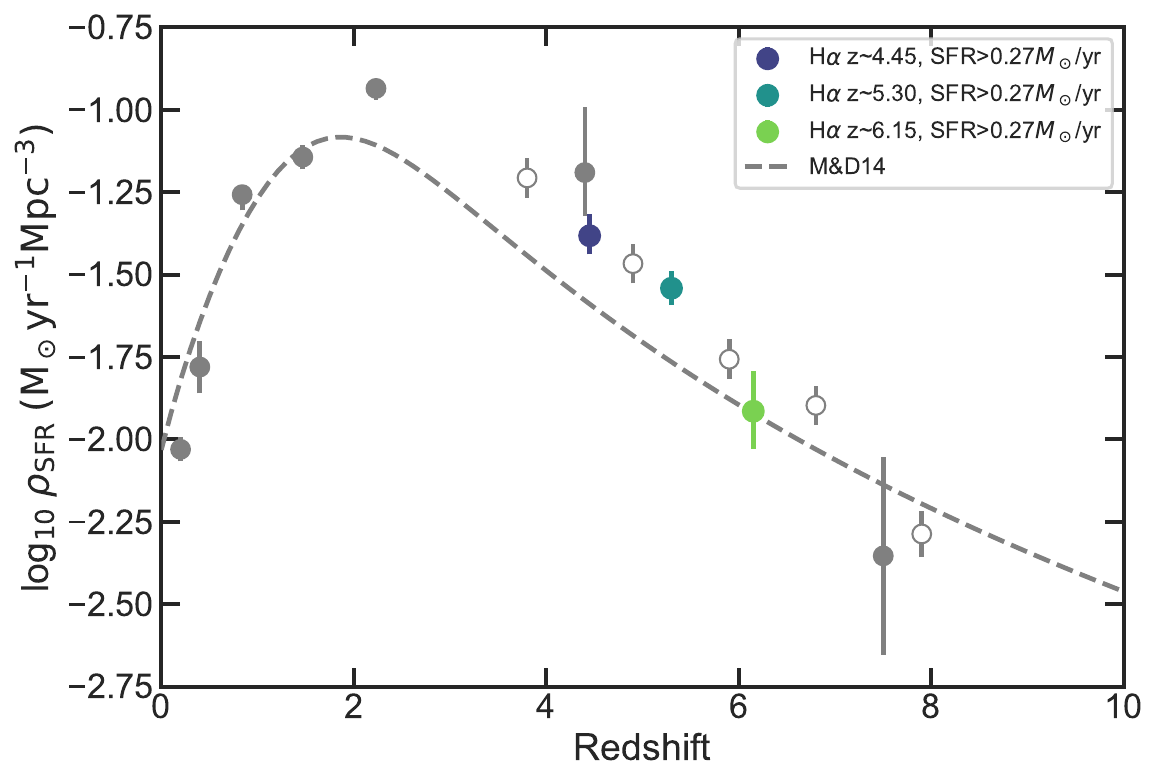}
\caption{Cosmic evolution of the SFRD: our values compared to previous results: \citet{Stroe&Sobral15} at $z\sim0.2$; \citet{Sobral+13} at $z\sim$0.4, 0.84, 1,47, and 2.23; \citet{Terao+22} at $z\sim2.3$; \citet{Bollo23} at $z\sim4.4$; and \citet{Rinaldi+23} at $z\sim7.5$. Filled circles represent values obtained from H$\alpha$ luminosity functions, while open circles are from \citet{Bouwens+15} UV luminosity functions. The dotted line is the fit by \citet{MD14}. All literature results were scaled to match our H$\alpha$ conversion factor.}
\label{SFRD}
\end{figure*}

Fig. \ref{SFRD} shows the SFRD values at various redshifts obtained from integrating the H$\alpha$ luminosity functions down to SFR$_l=0.27M_\odot\,\textrm{yr}^{-1}$. For comparison, we also included values obtained from UV measurements by \citet{Bouwens+15}. This provides us with a view of the star formation history of the universe at redshifts $0\leq z\leq8$. We find that both H$\alpha$ and UV measurements tend to lie $\sim0.2$ dex above the parametrization of \citet{MD14}, which was derived from UV and IR measurements prior to \citet{Bouwens+15} and integrating down to SFR$_l=0.03L^*$. This disagreement arises from the difference between integration limits. 

It is important to note that both H$\alpha$-derived and UV-derived SFRD measurements agree remarkably well with each other. Given that H$\alpha$ traces star formation at much shorter timescales than UV ($\sim10$Myr vs $\sim100$Myr), it is common that both tracers provide different SFR values for the same galaxy due to burstiness \citep[e.g.,][]{Cole23,Calabro24,Clarke24}. Additionally, they are subject to different dust attenuation.
Nevertheless, both SFRD measurements agreeing at the studied redshift range shows that these effects average out upon integration. 

\section{Summary and conclusions}
In this work, we presented a new catalog of H$\alpha$ emitters in the GOODS fields obtained using NIRCam/grism spectroscopy from \emph{JWST}'s FRESCO and CONGRESS surveys. These $1\,050$ sources entail a completeness-characterized census of line emitters at $3.8<z<6.6$ in GOODS-North and at $4.9<z<6.6$ in GOODS-South. With this catalog, we find a newly-discovered overdensity of 98 H$\alpha$ emitters in GOODS-North at $z\sim4.4$ and we confirm the overdensities reported by \citet{Helton+23}. We also find five new broad H$\alpha$ emitters and confirm those reported as LRDs in \citet{LRDs}.

We computed the UV LFs and confirmed the photometry-based results from \citet{Bouwens21} that were based on Lyman break selections. 
We computed the first H$\alpha$ LFs at $z>3$
based purely on spectroscopic data. We also derived the H$\alpha$ LFs from different simulations, finding that our observed estimates are consistent with the predicted values. By comparing our observed results to previous H$\alpha$ LFs over a range of redshifts, we find that both the bright and faint ends evolve from the local universe by increasing with increasing redshift up to a maximum at $z\sim2-3$, to then decrease back. This picture is consistent with the SFRD evolution described by \citet{MD14}, which we confirmed by obtaining three new values at redshifts $4.45$, $5.3$, and $6.15$ from the integrated dust-corrected H$\alpha$ luminosity functions. Our SFRD values agree with those derived from UV measurements, despite H$\alpha$ and UV tracing star formation at different timescales. 

While fitting Schechter functions to the observed H$\alpha$ luminosity functions, we find a large uncertainty in the faint-end slope, $\alpha$, which is also degenerate with the characteristic number density, $\phi^*$. This suggests that it is necessary to conduct deeper observations using NIRSpec to probe the faint end of the H$\alpha$ luminosity function (as is being done with, e.g., the ALT survey). At the same time, COSMOS-3D \citep{cosmos3d} and parallel programs will increase the observed area and find the most massive galaxies, probing the brighter end of the luminosity functions. 
Our work demonstrates the power of \emph{JWST} NIRCam/grism to efficiently obtain complete emission-line-selected galaxy samples in a limited observation time (only $\sim2$h on source for FRESCO). In turn, these galaxy samples provide completely new constraints on galaxy formation models.

\begin{acknowledgements}
This work is based on observations made with the NASA/ESA/CSA James Webb Space Telescope. The data were obtained from the Mikulski Archive for Space Telescopes at the Space Telescope Science Institute, which is operated by the Association of Universities for Research in Astronomy, Inc., under NASA contract NAS 5-03127 for JWST. These observations are associated with program Nos. 1895 and 3577. The authors sincerely thank the CONGRESS team (PIs: Egami \& Sun) for developing their observing program with a zero-exclusive-access period.
We thank Aswin Vijayan and Harley Katz for their help in analyzing the simulation data from FLARES and SPHINX. 
This work has received funding from the Swiss State Secretariat for Education, Research, and Innovation (SERI) under contract number MB22.00072, as well as from the Swiss National Science Foundation (SNSF) through project grant 200020\_207349. The Cosmic Dawn Center (DAWN) is funded by the Danish National Research Foundation under grant DNRF140. 
Support for program \#1895 was provided by NASA through a grant
from the Space Telescope Science Institute, which is operated by the
Association of Universities for Research in Astronomy, Inc., under
NASA contract NAS 5-03127. Support for this work for RPN was provided by NASA through the NASA Hubble Fellowship grant HST-HF2-51515.001-A awarded by the Space Telescope Science Institute, which is operated by the Association of Universities for Research in Astronomy, Incorporated, under NASA contract NAS5-26555. 
MS acknowledges support from the European Research Commission Consolidator Grant 101088789 (SFEER), from the CIDEGENT/2021/059 grant by Generalitat Valenciana, and from project PID2023-149420NB-I00 funded by MICIU/AEI/10.13039/501100011033 and by ERDF/EU. 
\end{acknowledgements}

\bibliographystyle{aa}
\bibliography{references}

\begin{thebibliography}{65}
\expandafter\ifx\csname natexlab\endcsname\relax\def\natexlab#1{#1}\fi

\bibitem[{{Bertin} \& {Arnouts}(1996)}]{Bertin1996}
{Bertin}, E. \& {Arnouts}, S. 1996, \aaps, 117, 393

\bibitem[{{Bollo} {et~al.}(2023){Bollo}, {Gonz{\'a}lez}, {Stefanon}, {Oesch}, {Bouwens}, {Smit}, {Illingworth}, \& {Labb{\'e}}}]{Bollo23}
{Bollo}, V., {Gonz{\'a}lez}, V., {Stefanon}, M., {et~al.} 2023, \apj, 946, 117

\bibitem[{{Bouwens} {et~al.}(2012{\natexlab{a}}){Bouwens}, {Illingworth}, {Oesch}, {Franx}, {Labb{\'e}}, {Trenti}, {van Dokkum}, {Carollo}, {Gonz{\'a}lez}, {Smit}, \& {Magee}}]{Bouwens12b}
{Bouwens}, R.~J., {Illingworth}, G.~D., {Oesch}, P.~A., {et~al.} 2012{\natexlab{a}}, \apj, 754, 83

\bibitem[{{Bouwens} {et~al.}(2015){Bouwens}, {Illingworth}, {Oesch}, {Trenti}, {Labb{\'e}}, {Bradley}, {Carollo}, {van Dokkum}, {Gonzalez}, {Holwerda}, {Franx}, {Spitler}, {Smit}, \& {Magee}}]{Bouwens+15}
{Bouwens}, R.~J., {Illingworth}, G.~D., {Oesch}, P.~A., {et~al.} 2015, \apj, 803, 34

\bibitem[{{Bouwens} {et~al.}(2012{\natexlab{b}}){Bouwens}, {Illingworth}, {Oesch}, {Trenti}, {Labb{\'e}}, {Franx}, {Stiavelli}, {Carollo}, {van Dokkum}, \& {Magee}}]{Bouwens12a}
{Bouwens}, R.~J., {Illingworth}, G.~D., {Oesch}, P.~A., {et~al.} 2012{\natexlab{b}}, \apjl, 752, L5

\bibitem[{{Bouwens} {et~al.}(2021){Bouwens}, {Oesch}, {Stefanon}, {Illingworth}, {Labb{\'e}}, {Reddy}, {Atek}, {Montes}, {Naidu}, {Nanayakkara}, {Nelson}, \& {Wilkins}}]{Bouwens21}
{Bouwens}, R.~J., {Oesch}, P.~A., {Stefanon}, M., {et~al.} 2021, \aj, 162, 47

\bibitem[{{Calabr{\`o}} {et~al.}(2024){Calabr{\`o}}, {Pentericci}, {Santini}, {Ferrara}, {Llerena}, {Mascia}, {Napolitano}, {Yung}, {Bisigello}, {Castellano}, {Cleri}, {Dekel}, {Dickinson}, {Franco}, {Giavalisco}, {Hirschmann}, {Holwerda}, {Koekemoer}, {Lucas}, {Pacucci}, {Pirzkal}, {Roberts-Borsani}, {Seill{\'e}}, {Tacchella}, {Wilkins}, {Amor{\'\i}n}, {Arrabal Haro}, {Bagley}, {Finkelstein}, {Kartaltepe}, \& {Papovich}}]{Calabro24}
{Calabr{\`o}}, A., {Pentericci}, L., {Santini}, P., {et~al.} 2024, \aap, 690, A290

\bibitem[{{Calzetti} {et~al.}(2000){Calzetti}, {Armus}, {Bohlin}, {Kinney}, {Koornneef}, \& {Storchi-Bergmann}}]{Calzetti+00}
{Calzetti}, D., {Armus}, L., {Bohlin}, R.~C., {et~al.} 2000, \apj, 533, 682

\bibitem[{{Carnall} {et~al.}(2018){Carnall}, {McLure}, {Dunlop}, \& {Dav{\'e}}}]{Bagpipes}
{Carnall}, A.~C., {McLure}, R.~J., {Dunlop}, J.~S., \& {Dav{\'e}}, R. 2018, \mnras, 480, 4379

\bibitem[{{Chabrier}(2003)}]{Chabrier+03}
{Chabrier}, G. 2003, \pasp, 115, 763

\bibitem[{{Chatzikos} {et~al.}(2023){Chatzikos}, {Bianchi}, {Camilloni}, {Chakraborty}, {Gunasekera}, {Guzm{\'a}n}, {Milby}, {Sarkar}, {Shaw}, {van Hoof}, \& {Ferland}}]{Chetzikos23}
{Chatzikos}, M., {Bianchi}, S., {Camilloni}, F., {et~al.} 2023, \rmxaa, 59, 327

\bibitem[{{Clarke} {et~al.}(2024){Clarke}, {Shapley}, {Sanders}, {Topping}, {Brammer}, {Bento}, {Reddy}, \& {Kehoe}}]{Clarke24}
{Clarke}, L., {Shapley}, A.~E., {Sanders}, R.~L., {et~al.} 2024, \apj, 977, 133

\bibitem[{{Cole} {et~al.}(2023){Cole}, {Papovich}, {Finkelstein}, {Bagley}, {Dickinson}, {Iyer}, {Yung}, {Ciesla}, {Amorin}, {Arrabal Haro}, {Bhatawdekar}, {Calabro}, {Cleri}, {de la Vega}, {Dekel}, {Endsley}, {Gawiser}, {Giavalisco}, {Hathi}, {Hirschmann}, {Holwerda}, {Kartaltepe}, {Koekemoer}, {Lucas}, {Mascia}, {Mobasher}, {Perez-Gonzalez}, {Rodighiero}, {Ronayne}, {Tachhella}, {Weiner}, \& {Wilkins}}]{Cole23}
{Cole}, J.~W., {Papovich}, C., {Finkelstein}, S.~L., {et~al.} 2023, arXiv e-prints, arXiv:2312.10152

\bibitem[{{Cucciati} {et~al.}(2012){Cucciati}, {Tresse}, {Ilbert}, {Le F{\`e}vre}, {Garilli}, {Le Brun}, {Cassata}, {Franzetti}, {Maccagni}, {Scodeggio}, {Zucca}, {Zamorani}, {Bardelli}, {Bolzonella}, {Bielby}, {McCracken}, {Zanichelli}, \& {Vergani}}]{Cucciati12}
{Cucciati}, O., {Tresse}, L., {Ilbert}, O., {et~al.} 2012, \aap, 539, A31

\bibitem[{{Dahlen} {et~al.}(2007){Dahlen}, {Mobasher}, {Dickinson}, {Ferguson}, {Giavalisco}, {Kretchmer}, \& {Ravindranath}}]{Dahlen07}
{Dahlen}, T., {Mobasher}, B., {Dickinson}, M., {et~al.} 2007, \apj, 654, 172

\bibitem[{{D'Eugenio} {et~al.}(2024){D'Eugenio}, {Cameron}, {Scholtz}, {Carniani}, {Willott}, {Curtis-Lake}, {Bunker}, {Parlanti}, {Maiolino}, {Willmer}, {Jakobsen}, {Robertson}, {Johnson}, {Tacchella}, {Cargile}, {Rawle}, {Arribas}, {Chevallard}, {Curti}, {Egami}, {Eisenstein}, {Kumari}, {Looser}, {Rieke}, {Rodr{\'\i}guez Del Pino}, {Saxena}, {{\"U}bler}, {Venturi}, {Witstok}, {Baker}, {Bhatawdekar}, {Bonaventura}, {Boyett}, {Charlot}, {Danhaive}, {Hainline}, {Hausen}, {Helton}, {Ji}, {Ji}, {Jones}, {Joud{\v{z}}balis}, {Maseda}, {P{\'e}rez-Gonz{\'a}lez}, {Perna}, {Pusk{\'a}s}, {Shivaei}, {Silcock}, {Simmonds}, {Smit}, {Sun}, {Villanueva}, {Williams}, \& {Zhu}}]{JADES_DR3}
{D'Eugenio}, F., {Cameron}, A.~J., {Scholtz}, J., {et~al.} 2024, arXiv e-prints, arXiv:2404.06531

\bibitem[{{Egami} {et~al.}(2023){Egami}, {Sun}, {Alberts}, {Baum}, {Boyett}, {Bunker}, {Cameron}, {Carniani}, {Charlot}, {Chen}, {Chevallard}, {Curti}, {D'Eugenio}, {Danhaive}, {DeCoursey}, {Dudzeviciute}, {Eisenstein}, {Hainline}, {Helton}, {Ji}, {Johnson}, {Kumari}, {Looser}, {Lyu}, {Ma}, {Maiolino}, {Maseda}, {Nelson}, {Rawle}, {Rieke}, {Robertson}, {Sandles}, {Shivaei}, {Smit}, {Suess}, {Tacchella}, {Uebler}, {Whitler}, {Williams}, {Willmer}, {Willott}, {Witstok}, \& {de Graaff}}]{Congress}
{Egami}, E., {Sun}, F., {Alberts}, S., {et~al.} 2023, {Complete NIRCam Grism Redshift Survey (CONGRESS)}, JWST Proposal. Cycle 2, ID. \#3577

\bibitem[{{Eisenstein} {et~al.}(2023{\natexlab{a}}){Eisenstein}, {Johnson}, {Robertson}, {Tacchella}, {Hainline}, {Jakobsen}, {Maiolino}, {Bonaventura}, {Bunker}, {Cameron}, {Cargile}, {Curtis-Lake}, {Hausen}, {Pusk{\'a}s}, {Rieke}, {Sun}, {Willmer}, {Willott}, {Alberts}, {Arribas}, {Baker}, {Baum}, {Bhatawdekar}, {Carniani}, {Charlot}, {Chen}, {Chevallard}, {Curti}, {DeCoursey}, {D'Eugenio}, {de Graaff}, {Egami}, {Helton}, {Ji}, {Jones}, {Kumari}, {L{\"u}tzgendorf}, {Laseter}, {Looser}, {Lyu}, {Maseda}, {Nelson}, {Parlanti}, {Rauscher}, {Rawle}, {Rieke}, {Rix}, {Rujopakarn}, {Sandles}, {Saxena}, {Scholtz}, {Sharpe}, {Shivaei}, {Simmonds}, {Smit}, {Topping}, {{\"U}bler}, {Venturi}, {Williams}, {Witstok}, \& {Woodrum}}]{Eisenstein2023b}
{Eisenstein}, D.~J., {Johnson}, B.~D., {Robertson}, B., {et~al.} 2023{\natexlab{a}}, arXiv e-prints, arXiv:2310.12340

\bibitem[{{Eisenstein} {et~al.}(2023{\natexlab{b}}){Eisenstein}, {Willott}, {Alberts}, {Arribas}, {Bonaventura}, {Bunker}, {Cameron}, {Carniani}, {Charlot}, {Curtis-Lake}, {D'Eugenio}, {Endsley}, {Ferruit}, {Giardino}, {Hainline}, {Hausen}, {Jakobsen}, {Johnson}, {Maiolino}, {Rieke}, {Rieke}, {Rix}, {Robertson}, {Stark}, {Tacchella}, {Williams}, {Willmer}, {Baker}, {Baum}, {Bhatawdekar}, {Boyett}, {Chen}, {Chevallard}, {Circosta}, {Curti}, {Danhaive}, {DeCoursey}, {de Graaff}, {Dressler}, {Egami}, {Helton}, {Hviding}, {Ji}, {Jones}, {Kumari}, {L{\"u}tzgendorf}, {Laseter}, {Looser}, {Lyu}, {Maseda}, {Nelson}, {Parlanti}, {Perna}, {Pusk{\'a}s}, {Rawle}, {Rodr{\'\i}guez Del Pino}, {Sandles}, {Saxena}, {Scholtz}, {Sharpe}, {Shivaei}, {Silcock}, {Simmonds}, {Skarbinski}, {Smit}, {Stone}, {Suess}, {Sun}, {Tang}, {Topping}, {{\"U}bler}, {Villanueva}, {Wallace}, {Whitler}, {Witstok}, \& {Woodrum}}]{Eisenstein2023a}
{Eisenstein}, D.~J., {Willott}, C., {Alberts}, S., {et~al.} 2023{\natexlab{b}}, arXiv e-prints, arXiv:2306.02465

\bibitem[{{Erb} {et~al.}(2006){Erb}, {Steidel}, {Shapley}, {Pettini}, {Reddy}, \& {Adelberger}}]{Erb06}
{Erb}, D.~K., {Steidel}, C.~C., {Shapley}, A.~E., {et~al.} 2006, \apj, 647, 128

\bibitem[{{Ferland} {et~al.}(1998){Ferland}, {Korista}, {Verner}, {Ferguson}, {Kingdon}, \& {Verner}}]{ferland98}
{Ferland}, G.~J., {Korista}, K.~T., {Verner}, D.~A., {et~al.} 1998, \pasp, 110, 761

\bibitem[{{F{\"o}rster Schreiber} {et~al.}(2009){F{\"o}rster Schreiber}, {Genzel}, {Bouch{\'e}}, {Cresci}, {Davies}, {Buschkamp}, {Shapiro}, {Tacconi}, {Hicks}, {Genel}, {Shapley}, {Erb}, {Steidel}, {Lutz}, {Eisenhauer}, {Gillessen}, {Sternberg}, {Renzini}, {Cimatti}, {Daddi}, {Kurk}, {Lilly}, {Kong}, {Lehnert}, {Nesvadba}, {Verma}, {McCracken}, {Arimoto}, {Mignoli}, \& {Onodera}}]{Forster09}
{F{\"o}rster Schreiber}, N.~M., {Genzel}, R., {Bouch{\'e}}, N., {et~al.} 2009, \apj, 706, 1364

\bibitem[{{Gardner} {et~al.}(2006){Gardner}, {Mather}, {Clampin}, {Doyon}, {Greenhouse}, {Hammel}, {Hutchings}, {Jakobsen}, {Lilly}, {Long}, {Lunine}, {McCaughrean}, {Mountain}, {Nella}, {Rieke}, {Rieke}, {Rix}, {Smith}, {Sonneborn}, {Stiavelli}, {Stockman}, {Windhorst}, \& {Wright}}]{Gardner+06}
{Gardner}, J.~P., {Mather}, J.~C., {Clampin}, M., {et~al.} 2006, \ssr, 123, 485

\bibitem[{{Geach} {et~al.}(2008){Geach}, {Smail}, {Best}, {Kurk}, {Casali}, {Ivison}, \& {Coppin}}]{Geach08}
{Geach}, J.~E., {Smail}, I., {Best}, P.~N., {et~al.} 2008, \mnras, 388, 1473

\bibitem[{Giavalisco {et~al.}(2004)Giavalisco, Ferguson, Koekemoer, Dickinson, Alexander, Bauer, Bergeron, Biagetti, Brandt, Casertano, Cesarsky, Chatzichristou, Conselice, Cristiani, Costa, Dahlen, de~Mello, Eisenhardt, Erben, Fall, Fassnacht, Fosbury, Fruchter, Gardner, Grogin, Hook, Hornschemeier, Idzi, Jogee, Kretchmer, Laidler, Lee, Livio, Lucas, Madau, Mobasher, Moustakas, Nonino, Padovani, Papovich, Park, Ravindranath, Renzini, Richardson, Riess, Rosati, Schirmer, Schreier, Somerville, Spinrad, Stern, Stiavelli, Strolger, Urry, Vandame, Williams, \& Wolf}]{Giavalisco2004}
Giavalisco, M., Ferguson, H.~C., Koekemoer, A.~M., {et~al.} 2004, The Astrophysical Journal, 600, L93

\bibitem[{Grogin {et~al.}(2011)Grogin, Kocevski, Faber, Ferguson, Koekemoer, Riess, Acquaviva, Alexander, Almaini, Ashby, Barden, Bell, Bournaud, Brown, Caputi, Casertano, Cassata, Castellano, Challis, Chary, Cheung, Cirasuolo, Conselice, Cooray, Croton, Daddi, Dahlen, Davé, Mello, Dekel, Dickinson, Dolch, Donley, Dunlop, Dutton, Elbaz, Fazio, Filippenko, Finkelstein, Fontana, Gardner, Garnavich, Gawiser, Giavalisco, Grazian, Guo, Hathi, Häussler, Hopkins, Huang, Huang, Jha, Kartaltepe, Kirshner, Koo, Lai, Lee, Li, Lotz, Lucas, Madau, McCarthy, McGrath, McIntosh, McLure, Mobasher, Moustakas, Mozena, Nandra, Newman, Niemi, Noeske, Papovich, Pentericci, Pope, Primack, Rajan, Ravindranath, Reddy, Renzini, Rix, Robaina, Rodney, Rosario, Rosati, Salimbeni, Scarlata, Siana, Simard, Smidt, Somerville, Spinrad, Straughn, Strolger, Telford, Teplitz, Trump, Wel, Villforth, Wechsler, Weiner, Wiklind, Wild, Wilson, Wuyts, Yan, \& Yun}]{Grogin2011}
Grogin, N.~A., Kocevski, D.~D., Faber, S.~M., {et~al.} 2011, Astrophysical Journal, Supplement Series, 197, 2011

\bibitem[{{Hayes} {et~al.}(2010){Hayes}, {Schaerer}, \& {{\"O}stlin}}]{Hayes10}
{Hayes}, M., {Schaerer}, D., \& {{\"O}stlin}, G. 2010, \aap, 509, L5

\bibitem[{{Helton} {et~al.}(2024){Helton}, {Sun}, {Woodrum}, {Hainline}, {Willmer}, {Rieke}, {Rieke}, {Alberts}, {Eisenstein}, {Tacchella}, {Robertson}, {Johnson}, {Baker}, {Bhatawdekar}, {Bunker}, {Chen}, {Egami}, {Ji}, {Maiolino}, {Willott}, \& {Witstok}}]{Helton+23}
{Helton}, J.~M., {Sun}, F., {Woodrum}, C., {et~al.} 2024, \apj, 974, 41

\bibitem[{{Herard-Demanche} {et~al.}(2023){Herard-Demanche}, {Bouwens}, {Oesch}, {Naidu}, {Decarli}, {Nelson}, {Brammer}, {Weibel}, {Xiao}, {Stefanon}, {Walter}, {Matthee}, {Meyer}, {Wuyts}, {Reddy}, {Arrabal Haro}, {Dannerbauer}, {Shapley}, {Chisholm}, {van Dokkum}, {Labbe}, {Illingworth}, {Schaerer}, \& {Shivaei}}]{HerardDemanche24}
{Herard-Demanche}, T., {Bouwens}, R.~J., {Oesch}, P.~A., {et~al.} 2023, arXiv e-prints, arXiv:2309.04525

\bibitem[{{Kakiichi} {et~al.}(2024){Kakiichi}, {Egami}, {Fan}, {Lyu}, {Wang}, {Yang}, {Bechtel}, {Behroozi}, {Bosman}, {Cai}, {Champagne}, {Davies}, {De Rosa}, {Decarli}, {Eilers}, {Ellis}, {Endsley}, {Farina}, {Finkelstein}, {Fujimoto}, {Hennawi}, {Inoue}, {Jiang}, {Jin}, {Khusanova}, {Kirkpatrick}, {Kocevski}, {Kulkarni}, {Lee}, {Liu}, {Meyer}, {Ono}, {Onoue}, {Ouchi}, {Papovich}, {Satyavolu}, {Schindler}, {Sun}, {Tee}, {Vestergaard}, {Zhang}, \& {Zou}}]{cosmos3d}
{Kakiichi}, K., {Egami}, E., {Fan}, X., {et~al.} 2024, {COSMOS-3D: A Legacy Spectroscopic/Imaging Survey of the Early Universe}, JWST Proposal. Cycle 3, ID. \#5893

\bibitem[{{Kashino} {et~al.}(2013){Kashino}, {Silverman}, {Rodighiero}, {Renzini}, {Arimoto}, {Daddi}, {Lilly}, {Sanders}, {Kartaltepe}, {Zahid}, {Nagao}, {Sugiyama}, {Capak}, {Carollo}, {Chu}, {Hasinger}, {Ilbert}, {Kajisawa}, {Kewley}, {Koekemoer}, {Kova{\v{c}}}, {Le F{\`e}vre}, {Masters}, {McCracken}, {Onodera}, {Scoville}, {Strazzullo}, {Symeonidis}, \& {Taniguchi}}]{Kashino+13}
{Kashino}, D., {Silverman}, J.~D., {Rodighiero}, G., {et~al.} 2013, \apjl, 777, L8

\bibitem[{{Katz} {et~al.}(2023){Katz}, {Rosdahl}, {Kimm}, {Blaizot}, {Choustikov}, {Farcy}, {Garel}, {Haehnelt}, {Michel-Dansac}, \& {Ocvirk}}]{sphinx}
{Katz}, H., {Rosdahl}, J., {Kimm}, T., {et~al.} 2023, The Open Journal of Astrophysics, 6, 44

\bibitem[{{Kennicutt}(1998)}]{Kennicutt+98}
{Kennicutt}, Robert~C., J. 1998, \apj, 498, 541

\bibitem[{{Koekemoer} {et~al.}(2011){Koekemoer}, {Faber}, {Ferguson}, {Grogin}, {Kocevski}, {Koo}, {Lai}, {Lotz}, {Lucas}, {McGrath}, {Ogaz}, {Rajan}, {Riess}, {Rodney}, {Strolger}, {Casertano}, {Castellano}, {Dahlen}, {Dickinson}, {Dolch}, {Fontana}, {Giavalisco}, {Grazian}, {Guo}, {Hathi}, {Huang}, {van der Wel}, {Yan}, {Acquaviva}, {Alexander}, {Almaini}, {Ashby}, {Barden}, {Bell}, {Bournaud}, {Brown}, {Caputi}, {Cassata}, {Challis}, {Chary}, {Cheung}, {Cirasuolo}, {Conselice}, {Roshan Cooray}, {Croton}, {Daddi}, {Dav{\'e}}, {de Mello}, {de Ravel}, {Dekel}, {Donley}, {Dunlop}, {Dutton}, {Elbaz}, {Fazio}, {Filippenko}, {Finkelstein}, {Frazer}, {Gardner}, {Garnavich}, {Gawiser}, {Gruetzbauch}, {Hartley}, {H{\"a}ussler}, {Herrington}, {Hopkins}, {Huang}, {Jha}, {Johnson}, {Kartaltepe}, {Khostovan}, {Kirshner}, {Lani}, {Lee}, {Li}, {Madau}, {McCarthy}, {McIntosh}, {McLure}, {McPartland}, {Mobasher}, {Moreira}, {Mortlock}, {Moustakas}, {Mozena}, {Nandra}, {Newman}, {Nielsen}, {Niemi}, {Noeske}, {Papovich},
  {Pentericci}, {Pope}, {Primack}, {Ravindranath}, {Reddy}, {Renzini}, {Rix}, {Robaina}, {Rosario}, {Rosati}, {Salimbeni}, {Scarlata}, {Siana}, {Simard}, {Smidt}, {Snyder}, {Somerville}, {Spinrad}, {Straughn}, {Telford}, {Teplitz}, {Trump}, {Vargas}, {Villforth}, {Wagner}, {Wandro}, {Wechsler}, {Weiner}, {Wiklind}, {Wild}, {Wilson}, {Wuyts}, \& {Yun}}]{Koekemoer2011}
{Koekemoer}, A.~M., {Faber}, S.~M., {Ferguson}, H.~C., {et~al.} 2011, \apjs, 197, 36

\bibitem[{{Kriek} {et~al.}(2015){Kriek}, {Shapley}, {Reddy}, {Siana}, {Coil}, {Mobasher}, {Freeman}, {de Groot}, {Price}, {Sanders}, {Shivaei}, {Brammer}, {Momcheva}, {Skelton}, {van Dokkum}, {Whitaker}, {Aird}, {Azadi}, {Kassis}, {Bullock}, {Conroy}, {Dav{\'e}}, {Kere{\v{s}}}, \& {Krumholz}}]{Kriek+15}
{Kriek}, M., {Shapley}, A.~E., {Reddy}, N.~A., {et~al.} 2015, \apjs, 218, 15

\bibitem[{{Leethochawalit} {et~al.}(2022){Leethochawalit}, {Trenti}, {Morishita}, {Roberts-Borsani}, \& {Treu}}]{Leethochawalit2022}
{Leethochawalit}, N., {Trenti}, M., {Morishita}, T., {Roberts-Borsani}, G., \& {Treu}, T. 2022, \mnras, 509, 5836

\bibitem[{{Lovell} {et~al.}(2021){Lovell}, {Vijayan}, {Thomas}, {Wilkins}, {Barnes}, {Irodotou}, \& {Roper}}]{flares1}
{Lovell}, C.~C., {Vijayan}, A.~P., {Thomas}, P.~A., {et~al.} 2021, \mnras, 500, 2127

\bibitem[{{Madau} \& {Dickinson}(2014)}]{MD14}
{Madau}, P. \& {Dickinson}, M. 2014, \araa, 52, 415

\bibitem[{{Matharu} {et~al.}(2024){Matharu}, {Nelson}, {Brammer}, {Oesch}, {Allen}, {Shivaei}, {Naidu}, {Chisholm}, {Covelo-Paz}, {Fudamoto}, {Giovinazzo}, {Herard-Demanche}, {Kerutt}, {Kramarenko}, {Marchesini}, {Meyer}, {Prieto-Lyon}, {Reddy}, {Shuntov}, {Weibel}, {Wuyts}, \& {Xiao}}]{Matharu24}
{Matharu}, J., {Nelson}, E.~J., {Brammer}, G., {et~al.} 2024, \aap, 690, A64

\bibitem[{{Matthee} {et~al.}(2023){Matthee}, {Mackenzie}, {Simcoe}, {Kashino}, {Lilly}, {Bordoloi}, \& {Eilers}}]{Matthee+23}
{Matthee}, J., {Mackenzie}, R., {Simcoe}, R.~A., {et~al.} 2023, \apj, 950, 67

\bibitem[{{Matthee} {et~al.}(2024){Matthee}, {Naidu}, {Brammer}, {Chisholm}, {Eilers}, {Goulding}, {Greene}, {Kashino}, {Labbe}, {Lilly}, {Mackenzie}, {Oesch}, {Weibel}, {Wuyts}, {Xiao}, {Bordoloi}, {Bouwens}, {van Dokkum}, {Illingworth}, {Kramarenko}, {Maseda}, {Mason}, {Meyer}, {Nelson}, {Reddy}, {Shivaei}, {Simcoe}, \& {Yue}}]{LRDs}
{Matthee}, J., {Naidu}, R.~P., {Brammer}, G., {et~al.} 2024, \apj, 963, 129

\bibitem[{{Meyer} {et~al.}(2024){Meyer}, {Oesch}, {Giovinazzo}, {Weibel}, {Brammer}, {Matthee}, {Naidu}, {Bouwens}, {Chisholm}, {Covelo-Paz}, {Fudamoto}, {Maseda}, {Nelson}, {Shivaei}, {Xiao}, {Herard-Demanche}, {Illingworth}, {Kerutt}, {Kramarenko}, {Labbe}, {Leonova}, {Magee}, {Matharu}, {Prieto Lyon}, {Reddy}, {Schaerer}, {Shapley}, {Stefanon}, {Wozniak}, \& {Wuyts}}]{Meyer+24}
{Meyer}, R.~A., {Oesch}, P.~A., {Giovinazzo}, E., {et~al.} 2024, \mnras, 535, 1067

\bibitem[{{Nelson} {et~al.}(2024){Nelson}, {Brammer}, {Gim{\'e}nez-Arteaga}, {Oesch}, {Naidu}, {{\"U}bler}, {Matharu}, {Shapley}, {Whitaker}, {Wisnioski}, {F{\"o}rster Schreiber}, {Smit}, {van Dokkum}, {Chisholm}, {Endsley}, {Hartley}, {Gibson}, {Giovinazzo}, {Illingworth}, {Labbe}, {Maseda}, {Matthee}, {Covelo Paz}, {Price}, {Reddy}, {Shivaei}, {Weibel}, {Wuyts}, {Xiao}, {Alberts}, {Baker}, {Bunker}, {Cameron}, {Charlot}, {Eisenstein}, {de Graaff}, {Ji}, {Johnson}, {Jones}, {Maiolino}, {Robertson}, {Sandles}, {Suess}, {Tacchella}, {Williams}, \& {Witstok}}]{Nelson23}
{Nelson}, E., {Brammer}, G., {Gim{\'e}nez-Arteaga}, C., {et~al.} 2024, \apjl, 976, L27

\bibitem[{{Oesch} {et~al.}(2023){Oesch}, {Brammer}, {Naidu}, {Bouwens}, {Chisholm}, {Illingworth}, {Matthee}, {Nelson}, {Qin}, {Reddy}, {Shapley}, {Shivaei}, {van Dokkum}, {Weibel}, {Whitaker}, {Wuyts}, {Covelo-Paz}, {Endsley}, {Fudamoto}, {Giovinazzo}, {Herard-Demanche}, {Kerutt}, {Kramarenko}, {Labbe}, {Leonova}, {Lin}, {Magee}, {Marchesini}, {Maseda}, {Mason}, {Matharu}, {Meyer}, {Neufeld}, {Prieto Lyon}, {Schaerer}, {Sharma}, {Shuntov}, {Smit}, {Stefanon}, {Wyithe}, \& {Xiao}}]{Oesch23}
{Oesch}, P.~A., {Brammer}, G., {Naidu}, R.~P., {et~al.} 2023, \mnras, 525, 2864

\bibitem[{{Oke} \& {Gunn}(1983)}]{Oke&Gunn}
{Oke}, J.~B. \& {Gunn}, J.~E. 1983, \apj, 266, 713

\bibitem[{{Pirie} {et~al.}(2024){Pirie}, {Best}, {Duncan}, {McLeod}, {Cochrane}, {Clausen}, {Dunlop}, {Flury}, {Geach}, {Hale}, {Ibar}, {Kondapally}, {Li}, {Matthee}, {McLure}, {Ossa-Fuentes}, {Patrick}, {Smail}, {Sobral}, {Stephenson}, {Stott}, \& {Swinbank}}]{Pirie+24}
{Pirie}, C.~A., {Best}, P.~N., {Duncan}, K.~J., {et~al.} 2024, arXiv e-prints, arXiv:2410.11808

\bibitem[{{Reddy} \& {Steidel}(2009)}]{ReddySteidel09}
{Reddy}, N.~A. \& {Steidel}, C.~C. 2009, \apj, 692, 778

\bibitem[{{Rigby} {et~al.}(2023){Rigby}, {Perrin}, {McElwain}, {Kimble}, {Friedman}, {Lallo}, {Doyon}, {Feinberg}, {Ferruit}, {Glasse}, {Rieke}, {Rieke}, {Wright}, {Willott}, {Colon}, {Milam}, {Neff}, {Stark}, {Valenti}, {Abell}, {Abney}, {Abul-Huda}, {Acton}, {Adams}, {Adler}, {Aguilar}, {Ahmed}, {Albert}, {Alberts}, {Aldridge}, {Allen}, {Altenburg}, {{\'A}lvarez-M{\'a}rquez}, {Alves de Oliveira}, {Andersen}, {Anderson}, {Anderson}, {Argyriou}, {Armstrong}, {Arribas}, {Artigau}, {Arvai}, {Atkinson}, {Bacon}, {Bair}, {Banks}, {Barrientes}, {Barringer}, {Bartosik}, {Bast}, {Baudoz}, {Beatty}, {Bechtold}, {Beck}, {Bergeron}, {Bergkoetter}, {Bhatawdekar}, {Birkmann}, {Blazek}, {Blome}, {Boccaletti}, {B{\"o}ker}, {Boia}, {Bonaventura}, {Bond}, {Bosley}, {Boucarut}, {Bourque}, {Bouwman}, {Bower}, {Bowers}, {Boyer}, {Bradley}, {Brady}, {Braun}, {Breda}, {Bresnahan}, {Bright}, {Britt}, {Bromenschenkel}, {Brooks}, {Brooks}, {Brown}, {Brown}, {Brown}, {Bunker}, {Burger}, {Bushouse}, {Cale}, {Cameron}, {Cameron},
  {Canipe}, {Caplinger}, {Caputo}, {Cara}, {Carey}, {Carniani}, {Carrasquilla}, {Carruthers}, {Case}, {Catherine}, {Chance}, {Chapman}, {Charlot}, {Charlow}, {Chayer}, {Chen}, {Cherinka}, {Chichester}, {Chilton}, {Chonis}, {Clampin}, {Clark}, {Clark}, {Coe}, {Coleman}, {Comber}, {Comeau}, {Connolly}, {Cooper}, {Cooper}, {Coppock}, {Correnti}, {Cossou}, {Coulais}, {Coyle}, {Cracraft}, {Curti}, {Cuturic}, {Davis}, {Davis}, {Dean}, {DeLisa}, {deMeester}, {Dencheva}, {Dencheva}, {DePasquale}, {Deschenes}, {Hunor Detre}, {Diaz}, {Dicken}, {DiFelice}, {Dillman}, {Dixon}, {Doggett}, {Donaldson}, {Douglas}, {DuPrie}, {Dupuis}, {Durning}, {Easmin}, {Eck}, {Edeani}, {Egami}, {Ehrenwinkler}, {Eisenhamer}, {Eisenhower}, {Elie}, {Elliott}, {Elliott}, {Ellis}, {Engesser}, {Espinoza}, {Etienne}, {Etxaluze}, {Falini}, {Feeney}, {Ferry}, {Filippazzo}, {Fincham}, {Fix}, {Flagey}, {Florian}, {Flynn}, {Fontanella}, {Ford}, {Forshay}, {Fox}, {Franz}, {Fu}, {Fullerton}, {Galkin}, {Galyer}, {Garc{\'\i}a Mar{\'\i}n}, {Gardner},
  {Gardner}, {Garland}, {Garrett}, {Gasman}, {Gaspar}, {Gaudreau}, {Gauthier}, {Geers}, {Geithner}, {Gennaro}, {Giardino}, {Girard}, {Giuliano}, {Glassmire}, {Glauser}, {Glazer}, {Godfrey}, {Golimowski}, {Gollnitz}, {Gong}, {Gonzaga}, {Gordon}, {Gordon}, {Goudfrooij}, {Greene}, {Greenhouse}, {Grimaldi}, {Groebner}, {Grundy}, {Guillard}, {Gutman}, {Ha}, {Haderlein}, {Hagedorn}, {Hainline}, {Haley}, {Hami}, {Hamilton}, {Hammel}, {Hansen}, {Harkins}, {Harr}, {Hart}, {Hart}, {Hartig}, {Hashimoto}, {Haskins}, {Hathaway}, {Havey}, {Hayden}, {Hecht}, {Heller-Boyer}, {Henriques}, {Henry}, {Hermann}, {Hernandez}, {Hesman}, {Hicks}, {Hilbert}, {Hines}, {Hoffman}, {Holfeltz}, {Holler}, {Hoppa}, {Hott}, {Howard}, {Howard}, {Hunter}, {Hunter}, {Hurst}, {Husemann}, {Hustak}, {Ilinca Ignat}, {Illingworth}, {Irish}, {Jackson}, {Jahromi}, {Jakobsen}, {James}, {James}, {Januszewski}, {Jenkins}, {Jirdeh}, {Johnson}, {Johnson}, {Jones}, {Jones}, {Jones}, {Jones}, {Jordan}, {Jordan}, {Jurczyk}, {Jurling}, {Kaleida}, {Kalmanson},
  {Kammerer}, {Kang}, {Kao}, {Karakla}, {Kavanagh}, {Kelly}, {Kendrew}, {Kennedy}, {Kenny}, {Keski-kuha}, {Keyes}, {Kidwell}, {Kinzel}, {Kirk}, {Kirkpatrick}, {Kirshenblat}, {Klaassen}, {Knapp}, {Knight}, {Knollenberg}, {Koehler}, {Koekemoer}, {Kovacs}, {Kulp}, {Kumari}, {Kyprianou}, {La Massa}, {Labador}, {Labiano}, {Lagage}, {Lajoie}, {Lallo}, {Lam}, {Lamb}, {Lambros}, {Lampenfield}, {Langston}, {Larson}, {Law}, {Lawrence}, {Lee}, {Leisenring}, {Lepo}, {Leveille}, {Levenson}, {Levine}, {Levy}, {Lewis}, {Lewis}, {Libralato}, {Lightsey}, {Link}, {Liu}, {Lo}, {Lockwood}, {Logue}, {Long}, {Long}, {Loomis}, {Lopez-Caniego}, {Lorenzo Alvarez}, {Love-Pruitt}, {Lucy}, {Luetzgendorf}, {Maghami}, {Maiolino}, {Major}, {Malla}, {Malumuth}, {Manjavacas}, {Mannfolk}, {Marrione}, {Marston}, {Martel}, {Maschmann}, {Masci}, {Masciarelli}, {Maszkiewicz}, {Mather}, {McKenzie}, {McLean}, {McMaster}, {Melbourne}, {Mel{\'e}ndez}, {Menzel}, {Merz}, {Meyett}, {Meza}, {Miskey}, {Misselt}, {Moller}, {Morrison}, {Morse}, {Moseley},
  {Mosier}, {Mountain}, {Mueckay}, {Mueller}, {Mullally}, {Murphy}, {Murray}, {Murray}, {Mustelier}, {Muzerolle}, {Mycroft}, {Myers}, {Myrick}, {Nanavati}, {Nance}, {Nayak}, {Naylor}, {Nelan}, {Nickson}, {Nielson}, {Nieto-Santisteban}, {Nikolov}, {Noriega-Crespo}, {O'Shaughnessy}, {O'Sullivan}, {Ochs}, {Ogle}, {Oleszczuk}, {Olmsted}, {Osborne}, {Ottens}, {Owens}, {Pacifici}, {Pagan}, {Page}, {Park}, {Parrish}, {Patapis}, {Paul}, {Pauly}, {Pavlovsky}, {Pedder}, {Peek}, {Pena-Guerrero}, {Penanen}, {Perez}, {Perna}, {Perriello}, {Phillips}, {Pietraszkiewicz}, {Pinaud}, {Pirzkal}, {Pitman}, {Piwowar}, {Platais}, {Player}, {Plesha}, {Pollizi}, {Polster}, {Pontoppidan}, {Porterfield}, {Proffitt}, {Pueyo}, {Pulliam}, {Quirt}, {Quispe Neira}, {Ramos Alarcon}, {Ramsay}, {Rapp}, {Rapp}, {Rauscher}, {Ravindranath}, {Rawle}, {Regan}, {Reichard}, {Reis}, {Ressler}, {Rest}, {Reynolds}, {Rhue}, {Richon}, {Rickman}, {Ridgaway}, {Ritchie}, {Rix}, {Robberto}, {Robinson}, {Robinson}, {Robinson}, {Rock}, {Rodriguez}, {Rodriguez
  Del Pino}, {Roellig}, {Rohrbach}, {Roman}, {Romelfanger}, {Rose}, {Roteliuk}, {Roth}, {Rothwell}, {Rowlands}, {Roy}, {Royer}, {Royle}, {Rui}, {Rumler}, {Runnels}, {Russ}, {Rustamkulov}, {Ryden}, {Ryer}, {Sabata}, {Sabatke}, {Sabbi}, {Samuelson}, {Sapp}, {Sappington}, {Sargent}, {Sauer}, {Scheithauer}, {Schlawin}, {Schlitz}, {Schmitz}, {Schneider}, {Schreiber}, {Schulze}, {Schwab}, {Scott}, {Sembach}, {Shanahan}, {Shaughnessy}, {Shaw}, {Shawger}, {Shay}, {Sheehan}, {Shen}, {Sherman}, {Shiao}, {Shih}, {Shivaei}, {Sienkiewicz}, {Sing}, {Sirianni}, {Sivaramakrishnan}, {Skipper}, {Sloan}, {Slocum}, {Slowinski}, {Smith}, {Smith}, {Smith}, {Smith}, {Snyder}, {Soh}, {Sohn}, {Soto}, {Spencer}, {Stallcup}, {Stansberry}, {Starr}, {Starr}, {Stewart}, {Stiavelli}, {Straughn}, {Strickland}, {Stys}, {Summers}, {Sun}, {Sunnquist}, {Swade}, {Swam}, {Swaters}, {Swoish}, {Taylor}, {Taylor}, {Te Plate}, {Tea}, {Teague}, {Telfer}, {Temim}, {Thatte}, {Thompson}, {Thompson}, {Thomson}, {Tikkanen}, {Tippet}, {Todd}, {Toolan},
  {Tran}, {Trejo}, {Truong}, {Tsukamoto}, {Tustain}, {Tyra}, {Ubeda}, {Underwood}, {Uzzo}, {Van Campen}, {Vandal}, {Vandenbussche}, {Vila}, {Volk}, {Wahlgren}, {Waldman}, {Walker}, {Wander}, {Warfield}, {Warner}, {Wasiak}, {Watkins}, {Weaver}, {Weilert}, {Weiser}, {Weiss}, {Weissman}, {Welty}, {West}, {Wheate}, {Wheatley}, {Wheeler}, {White}, {Whiteaker}, {Whitehouse}, {Whiteleather}, {Whitman}, {Williams}, {Willmer}, {Willoughby}, {Wilson}, {Wirth}, {Wislowski}, {Wolf}, {Wolfe}, {Wolff}, {Workman}, {Wright}, {Wu}, {Wu}, {Wymer}, {Yates}, {Yeager}, {Yeates}, {Yerger}, {Yoon}, {Young}, {Yu}, {Zak}, {Zeidler}, {Zhou}, {Zielinski}, {Zincke}, \& {Zonak}}]{Rigby+23}
{Rigby}, J., {Perrin}, M., {McElwain}, M., {et~al.} 2023, \pasp, 135, 048001

\bibitem[{{Rinaldi} {et~al.}(2023){Rinaldi}, {Caputi}, {Costantin}, {Gillman}, {Iani}, {P{\'e}rez-Gonz{\'a}lez}, {{\"O}stlin}, {Colina}, {Greve}, {Noorgard-Nielsen}, {Wright}, {Alonso-Herrero}, {{\'A}lvarez-M{\'a}rquez}, {Eckart}, {Garc{\'\i}a-Mar{\'\i}n}, {Hjorth}, {Ilbert}, {Kendrew}, {Labiano}, {Le F{\`e}vre}, {Pye}, {Tikkanen}, {Walter}, {van der Werf}, {Ward}, {Annunziatella}, {Azzollini}, {Bik}, {Boogaard}, {Bosman}, {Crespo G{\'o}mez}, {Jermann}, {Langeroodi}, {Melinder}, {Meyer}, {Moutard}, {Peissker}, {Topinka}, {van Dishoeck}, {G{\"u}del}, {Henning}, {Lagage}, {Ray}, {Vandenbussche}, {Waelkens}, {Navarro-Carrera}, \& {Kokorev}}]{Rinaldi+23}
{Rinaldi}, P., {Caputi}, K.~I., {Costantin}, L., {et~al.} 2023, \apj, 952, 143

\bibitem[{{Robotham} \& {Driver}(2011)}]{RobothamDriver11}
{Robotham}, A.~S.~G. \& {Driver}, S.~P. 2011, \mnras, 413, 2570

\bibitem[{Rosdahl {et~al.}(2022)Rosdahl, Blaizot, Katz, Kimm, Garel, Haehnelt, Keating, Martin-Alvarez, Michel-Dansac, \& Ocvirk}]{Rosdahl_22}
Rosdahl, J., Blaizot, J., Katz, H., {et~al.} 2022, Monthly Notices of the Royal Astronomical Society, 515, 2386–2414

\bibitem[{{Schenker} {et~al.}(2013){Schenker}, {Robertson}, {Ellis}, {Ono}, {McLure}, {Dunlop}, {Koekemoer}, {Bowler}, {Ouchi}, {Curtis-Lake}, {Rogers}, {Schneider}, {Charlot}, {Stark}, {Furlanetto}, \& {Cirasuolo}}]{Schenker13}
{Schenker}, M.~A., {Robertson}, B.~E., {Ellis}, R.~S., {et~al.} 2013, \apj, 768, 196

\bibitem[{{Schiminovich} {et~al.}(2005){Schiminovich}, {Ilbert}, {Arnouts}, {Milliard}, {Tresse}, {Le F{\`e}vre}, {Treyer}, {Wyder}, {Budav{\'a}ri}, {Zucca}, {Zamorani}, {Martin}, {Adami}, {Arnaboldi}, {Bardelli}, {Barlow}, {Bianchi}, {Bolzonella}, {Bottini}, {Byun}, {Cappi}, {Contini}, {Charlot}, {Donas}, {Forster}, {Foucaud}, {Franzetti}, {Friedman}, {Garilli}, {Gavignaud}, {Guzzo}, {Heckman}, {Hoopes}, {Iovino}, {Jelinsky}, {Le Brun}, {Lee}, {Maccagni}, {Madore}, {Malina}, {Marano}, {Marinoni}, {McCracken}, {Mazure}, {Meneux}, {Morrissey}, {Neff}, {Paltani}, {Pell{\`o}}, {Picat}, {Pollo}, {Pozzetti}, {Radovich}, {Rich}, {Scaramella}, {Scodeggio}, {Seibert}, {Siegmund}, {Small}, {Szalay}, {Vettolani}, {Welsh}, {Xu}, \& {Zanichelli}}]{Schiminovich05}
{Schiminovich}, D., {Ilbert}, O., {Arnouts}, S., {et~al.} 2005, \apjl, 619, L47

\bibitem[{{Schmidt}(1968)}]{Schmidt+68}
{Schmidt}, M. 1968, \apj, 151, 393

\bibitem[{{Smit} {et~al.}(2016){Smit}, {Bouwens}, {Labb{\'e}}, {Franx}, {Wilkins}, \& {Oesch}}]{Smit+16}
{Smit}, R., {Bouwens}, R.~J., {Labb{\'e}}, I., {et~al.} 2016, \apj, 833, 254

\bibitem[{{Sobral} {et~al.}(2013){Sobral}, {Smail}, {Best}, {Geach}, {Matsuda}, {Stott}, {Cirasuolo}, \& {Kurk}}]{Sobral+13}
{Sobral}, D., {Smail}, I., {Best}, P.~N., {et~al.} 2013, \mnras, 428, 1128

\bibitem[{{Steidel} {et~al.}(2014){Steidel}, {Rudie}, {Strom}, {Pettini}, {Reddy}, {Shapley}, {Trainor}, {Erb}, {Turner}, {Konidaris}, {Kulas}, {Mace}, {Matthews}, \& {McLean}}]{Steidel+14}
{Steidel}, C.~C., {Rudie}, G.~C., {Strom}, A.~L., {et~al.} 2014, \apj, 795, 165

\bibitem[{{Stroe} \& {Sobral}(2015)}]{Stroe&Sobral15}
{Stroe}, A. \& {Sobral}, D. 2015, \mnras, 453, 242

\bibitem[{{Sun} {et~al.}(2023){Sun}, {Egami}, {Pirzkal}, {Rieke}, {Baum}, {Boyer}, {Boyett}, {Bunker}, {Cameron}, {Curti}, {Eisenstein}, {Gennaro}, {Greene}, {Jaffe}, {Kelly}, {Koekemoer}, {Kumari}, {Maiolino}, {Maseda}, {Perna}, {Rest}, {Robertson}, {Schlawin}, {Smit}, {Stansberry}, {Sunnquist}, {Tacchella}, {Williams}, \& {Willmer}}]{Sun+23}
{Sun}, F., {Egami}, E., {Pirzkal}, N., {et~al.} 2023, \apj, 953, 53

\bibitem[{{Terao} {et~al.}(2022){Terao}, {Spitler}, {Motohara}, \& {Chen}}]{Terao+22}
{Terao}, Y., {Spitler}, L.~R., {Motohara}, K., \& {Chen}, N. 2022, \apj, 941, 70

\bibitem[{{Trenti} \& {Stiavelli}(2008)}]{Trenti+08}
{Trenti}, M. \& {Stiavelli}, M. 2008, \apj, 676, 767

\bibitem[{{Vijayan} {et~al.}(2021){Vijayan}, {Lovell}, {Wilkins}, {Thomas}, {Barnes}, {Irodotou}, {Kuusisto}, \& {Roper}}]{flares2}
{Vijayan}, A.~P., {Lovell}, C.~C., {Wilkins}, S.~M., {et~al.} 2021, \mnras, 501, 3289

\bibitem[{{Weibel} {et~al.}(2024){Weibel}, {Oesch}, {Barrufet}, {Gottumukkala}, {Ellis}, {Santini}, {Weaver}, {Allen}, {Bouwens}, {Bowler}, {Brammer}, {Carnall}, {Cullen}, {Dayal}, {Dickinson}, {Donnan}, {Dunlop}, {Giavalisco}, {Grogin}, {Illingworth}, {Koekemoer}, {Labbe}, {Marchesini}, {McLeod}, {McLure}, {Naidu}, {P{\'e}rez-Gonz{\'a}lez}, {Shuntov}, {Stefanon}, {Toft}, \& {Xiao}}]{Weibel2024}
{Weibel}, A., {Oesch}, P.~A., {Barrufet}, L., {et~al.} 2024, \mnras, 533, 1808

\bibitem[{{Williams} {et~al.}(2023){Williams}, {Tacchella}, {Maseda}, {Robertson}, {Johnson}, {Willott}, {Eisenstein}, {Willmer}, {Ji}, {Hainline}, {Helton}, {Alberts}, {Baum}, {Bhatawdekar}, {Boyett}, {Bunker}, {Carniani}, {Charlot}, {Chevallard}, {Curtis-Lake}, {de Graaff}, {Egami}, {Franx}, {Kumari}, {Maiolino}, {Nelson}, {Rieke}, {Sandles}, {Shivaei}, {Simmonds}, {Smit}, {Suess}, {Sun}, {{\"U}bler}, \& {Witstok}}]{Williams2023}
{Williams}, C.~C., {Tacchella}, S., {Maseda}, M.~V., {et~al.} 2023, \apjs, 268, 64

\bibitem[{{Wyder} {et~al.}(2005){Wyder}, {Treyer}, {Milliard}, {Schiminovich}, {Arnouts}, {Budav{\'a}ri}, {Barlow}, {Bianchi}, {Byun}, {Donas}, {Forster}, {Friedman}, {Heckman}, {Jelinsky}, {Lee}, {Madore}, {Malina}, {Martin}, {Morrissey}, {Neff}, {Rich}, {Siegmund}, {Small}, {Szalay}, \& {Welsh}}]{Wyder05}
{Wyder}, T.~K., {Treyer}, M.~A., {Milliard}, B., {et~al.} 2005, \apjl, 619, L15

\end{thebibliography}

\begin{appendix}
\section{Catalog of H$\alpha$ emitters}
We present in table \ref{tab:cat} a fragment of our H$\alpha$ emitter catalog. The full catalog can be found at \url{https://github.com/astroalba/fresco}.

\begin{table*}
    \centering
\renewcommand{\arraystretch}{1.3} 
    \begin{tabular}{cccccccc}
    \hline
     ID & RA & DEC & $z_\textrm{spec}$ & $f_{\textrm{H}\alpha}$ &$M_\mathrm{UV}$& $f_\textrm{det}$ & q  \\
     & deg & deg & &  $10^{-18}\textrm{erg}\,\textrm{s}^{-1}\textrm{cm}^{-2}$ & & nJy & \\
     \hline
    GN-70 & 189.22095 & 62.33179 & 6.572 & $5.74\pm 0.34$ & $-20.85\pm0.05$ & $142.1\pm 8.1$ & 3.0\\
    GN-217 & 189.23841 & 62.32711 & 6.304 & $7.79\pm 0.42$ & $-21.12\pm0.06$ & $227.5\pm 8.4$ & 3.0\\
    GN-241 & 189.21517 & 62.32686 & 5.671 & $2.99\pm 0.21$ & $-20.14\pm0.11$ & $82.7\pm 6.5$ & 3.0\\
    GN-809 & 189.25302 & 62.31617 & 5.488 & $5.81\pm 0.29$ & $-19.73\pm0.09$ & $132.2\pm 5.4$ & 3.0\\
    \hline
    \end{tabular}
    \caption{Fragment of the full H$\alpha$ emitter catalog. IDs refer to the internal FRESCO data release v7.3. The detection flux is a combination of the F444W and F210M fluxes.}
    \label{tab:cat}
\end{table*}

\section{Repeated sources}
Within our catalog of $1\,050$ H$\alpha$ emitters, we find seven sources to show H$\alpha$ emission in both FRESCO and CONGRESS surveys, two of which are multi-component. This overlap is only technically possible at $4.86<z<5.15$, which in both surveys correspond to the edge of the grism, and thus the sensitivity at this redshift range is much lower than at the center of the grism. In fact, our seven overlapping sources are in the subrange $4.91<z<5.03$, where there are only 17 confirmed sources in FRESCO and 21 in CONGRESS. Moreover, the overlapping sources are the brighter emitters in this redshift range, and only the fainter ones are not retrieved in the other survey, respectively. 

The seven sources that we retrieve from both surveys show the same grism redshift, within a scatter of $0.05\times(1+z)$, and the same flux without an offset, within $0.06\sigma$.

\section{Flux calibration}
\label{sec:jades}
\begin{figure*}
    \centering
    \includegraphics[width=0.49\linewidth]{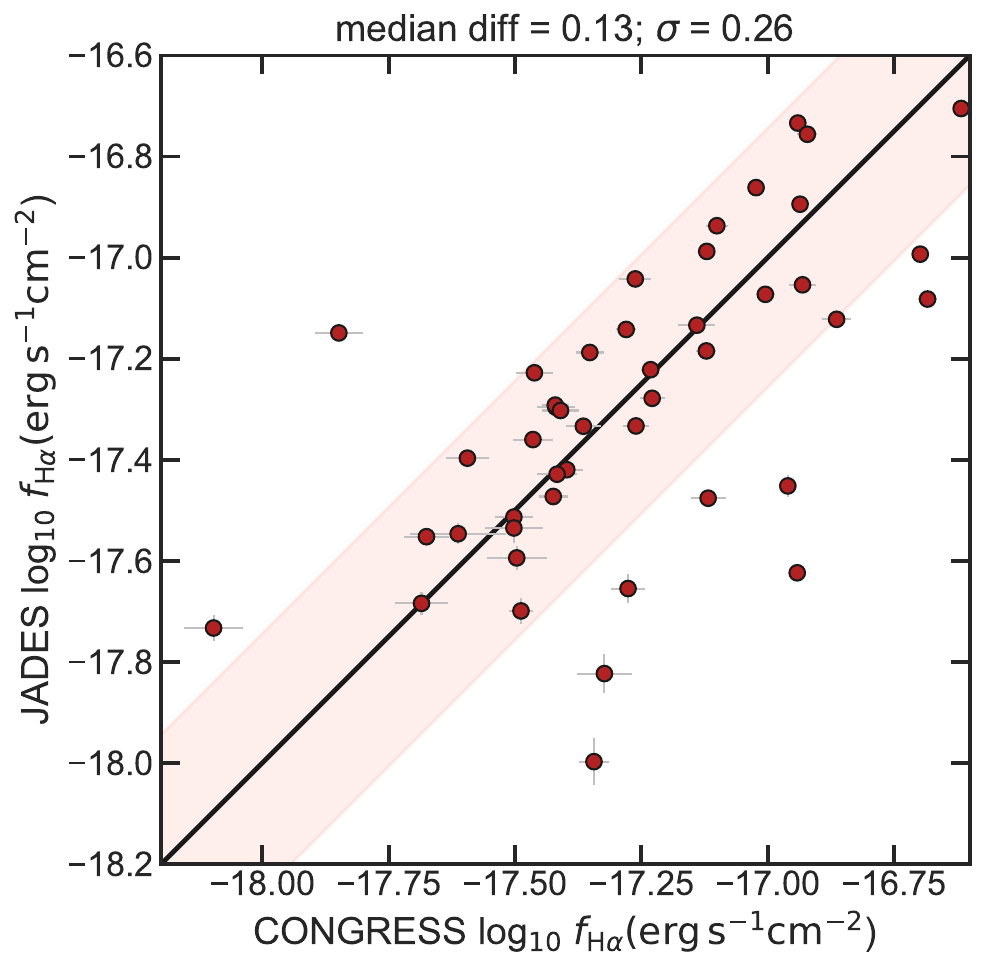}
    \includegraphics[width=0.49\linewidth]{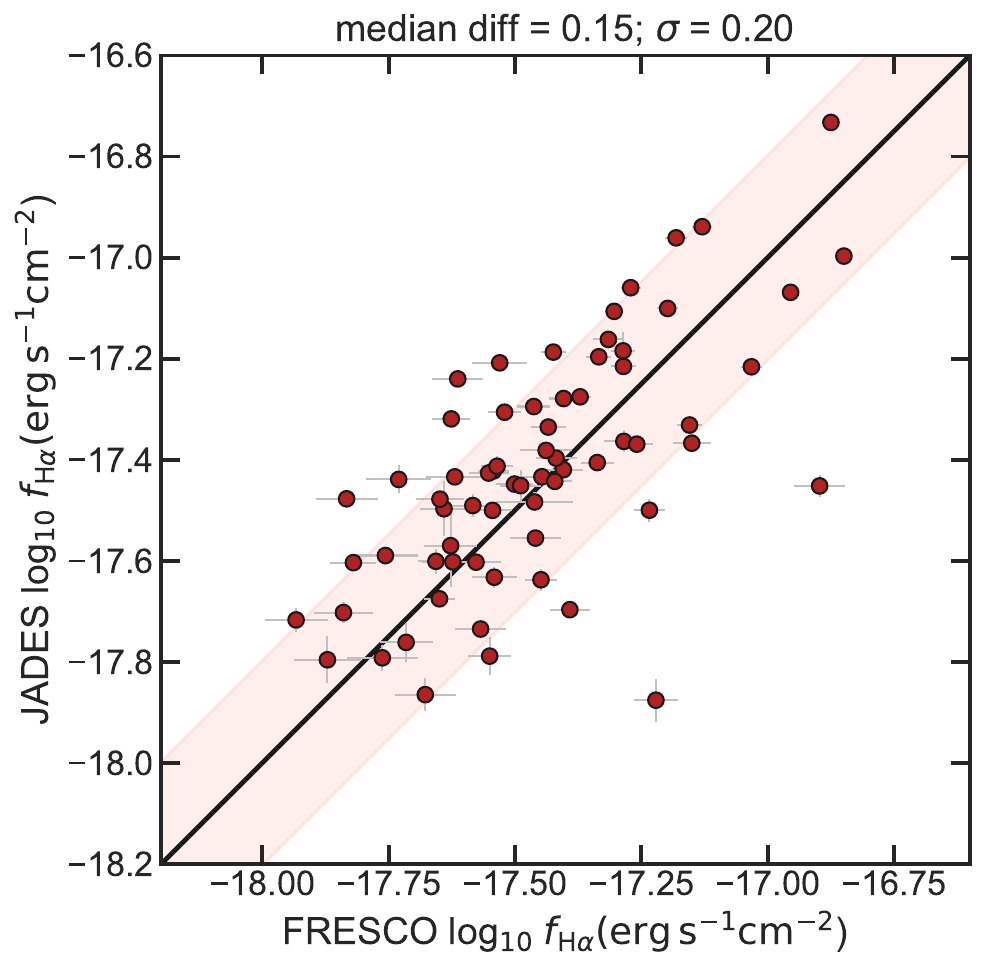}
    \caption{Comparison between our H$\alpha$ fluxes, obtained with NIRCam/grism, and the H$\alpha$ fluxes in JADES DR3 \citep{JADES_DR3}, obtained with NIRSpec prism. The solid lines are the 1:1 relations, while the shaded regions are the $1\sigma$ deviations. The median offsets written on top are calculated as} $\log(f_\textrm{grism}/f_\textrm{JADES})$.
    \label{usvsjades}
     \end{figure*}

In Fig. \ref{usvsjades} we compare our H$\alpha$ measurements to those obtained by JADES DR3 \citep{JADES_DR3} for the 123 sources present in both catalogs. Our measurements were obtained with NIRCam/grism spectroscopy, while JADES DR3 used NIRSpec G140M/G235M/G395M grating and prism
spectroscopy. We find a median offset on the H$\alpha$ fluxes of $0.13-0.15$ dex (calculated as $\log(f_\textrm{grism}/f_\textrm{JADES})$), which likely arises from an undercorrection on the NIRSpec fluxes. The scatter between the two measurements is likely due to calibration differences between NIRSpec and NIRCam/grism. We find that both calibrations are consistent within $0.26(0.20)\sigma$. 

\section{Broad H$\alpha$ emitters}
\label{sec:LRDs}
\begin{figure*}
    \centering
    \includegraphics[width=\linewidth]{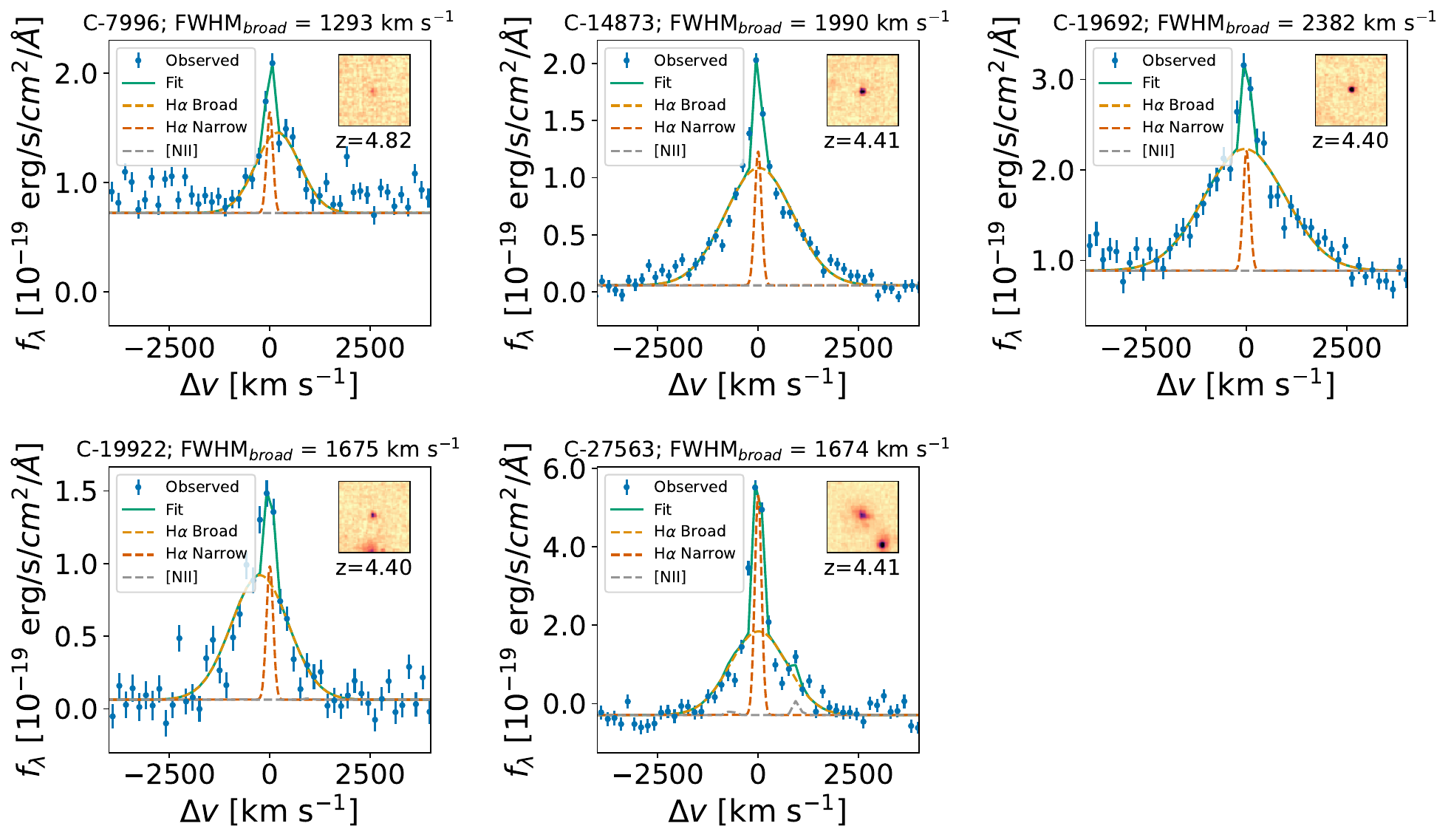}
    \caption{1D-spectra and F444W stamps of the five broad H$\alpha$ emitters in CONGRESS identified in this work. Filled circles represent the observed flux values of the spectra, solid lines are the multicomponent Gaussian fits (continuum+H$\alpha$+[\ion{N}{ii}]), and dashed lines are the H$\alpha$ broad, narrow, and [\ion{N}{ii}] components of these fits. The spectra are centered on the redshift of the narrow component.}
    \label{Fig:broad}
     \end{figure*}

Upon visual inspection, we found 13 of our sources to show broad H$\alpha$ lines likely consistent with AGN activity: five in CONGRESS and eight in FRESCO. The eight FRESCO sources are those reported by \citet{LRDs}, where they have been classified as Little Red Dots, while the five sources in CONGRESS have not been reported before, and thus we performed an analysis to determine the FWHM of their broad emission. Following the procedure in \citet{LRDs}, we fit their emission-line profiles with a combination of narrow and broad components of H$\alpha$ and [\ion{N}{ii}]. We find the values reported in table \ref{tab:lrds}. Regarding the morphology, four out of the five broad H$\alpha$ emitters in CONGRESS are point-like sources, as shown on the stamps in Fig. \ref{Fig:broad}. 

\begin{table*}
    \centering
\renewcommand{\arraystretch}{1.3} 
    \begin{tabular}{ccccc}
    \hline
     ID & RA & DEC & $z_\textrm{spec}$ & FWHM\\
     & deg & deg & & $\textrm{km}\,\textrm{s}^{-1}$ \\
     \hline
    CONGRESS-7996 & 189.187148 & 62.272917 & 4.821 &$1293\pm178$\\
    CONGRESS-14873 & 189.161854 & 62.251068 & 4.409 &$1990\pm131$\\
    CONGRESS-19692 & 189.286517 & 62.238126 & 4.404 &$2383\pm210$\\
    CONGRESS-19922 & 189.305664 & 62.236941 & 4.402 &$1675\pm102$\\
    CONGRESS-27563 & 189.276132 & 62.214160 & 4.406 &$1674\pm143$\\
    FRESCO-GN-3549 & 189.179308 & 62.292535 & 5.359 &$1443\pm138$\\
    FRESCO-GN-5896 & 189.285526 & 62.280772 & 5.087 &$1503\pm185$\\
    FRESCO-GN-7827 & 189.072083 & 62.273428 & 5.145 &$1473\pm107$\\
    FRESCO-GN-9123 & 189.057063 & 62.268938 & 5.245 &$2206\pm171$\\
    FRESCO-GN-10745 & 189.344287 & 62.263365 & 5.242 &$2949\pm180$\\
    FRESCO-GN-16262 & 189.281029 & 62.247311 & 5.539 &$2556\pm100$\\
    FRESCO-GN-26821 & 189.300147 & 62.212065 & 5.225 &$1982\pm179$\\
    FRESCO-MultiGS-09 & 53.1385956 & -27.790274 & 5.483 &$1396\pm99$\\
    \hline
    \end{tabular}
    \caption{Coordinates, redshifts, and broad-component FWHM of the 13 broad H$\alpha$ emitters found in this work.}
    \label{tab:lrds}
\end{table*}

\end{appendix}

\end{document}